\documentclass[11pt,twoside,nohyper]{pallis}
\usepackage{rotating}
\usepackage{epsfig}
\usepackage{latexsym}
\usepackage{euscript}
\usepackage{amssymb}
\usepackage{fancyheda}
\usepackage{times}
\usepackage{caption}
\usepackage{slashed,cite}
\usepackage{upgreek}

\usepackage{amsmath}
\usepackage{amsfonts}
\usepackage{amsbsy}
\usepackage{amscd}
\usepackage{bbm}

\newcommand{\ftn}{\footnotesize}
\newcommand{\nsz}{\normalsize}
\newcommand{\ssz}{\scriptsize}

\newcommand{\TeV}{{\mbox{\rm TeV}}}

\newcommand{\GeV}{{\mbox{\rm GeV}}}

\newcommand{\eV}{{\mbox{\rm eV}}}

\newcommand{\bal}{\begin{align}}
\newcommand{\eal}{\end{align}}
\newcommand{\beqs}{\begin{subequations}}
\newcommand{\eeqs}{\end{subequations}}
\newcommand{\eec}{\end{center}}
\newcommand{\bec}{\begin{center}}

\newcommand{\eem}{\end{matrix}}
\newcommand{\bem}{\begin{matrix}}
\newcommand{\eeq}{\end{equation}}
\newcommand{\beq}{\begin{equation}}
\newcommand{\ba}{\begin{array}}
\newcommand{\ea}{\end{array}}
\newcommand{\bea}{\begin{eqnarray}}
\newcommand{\eea}{\end{eqnarray}}
\newcommand{\baq}{\begin{eqnarray}}
\newcommand{\eaq}{\end{eqnarray}}

\newcommand{\bl}{\ensuremath{U(1)_{B-L}}}

\newcommand{\sFref}[2]{Fig.~\ref{#1}-{\small\sf ({#2})}}
\newcommand{\sEref}[2]{Eq.~(\ref{#1}{{\small\sf  #2}})}

\newcommand{\Eref}[1]{Eq.~(\ref{#1})}
\newcommand{\Sref}[1]{Sec.~\ref{#1}}
\newcommand{\Srefs}[1]{Secs.~\ref{#1}}
\newcommand{\Fref}[1]{Fig.~\ref{#1}}
\newcommand{\Tref}[1]{Table~\ref{#1}}
\newcommand{\cref}[1]{Ref.~\cite{#1}}
\newcommand{\crefs}[1]{Refs.~\cite{#1}}
\newcommand{\etal}{{\it et al.\/}}
\renewcommand{\email}[1]{{{\sl e-mail address:}~{\tt #1}}}

\newcommand\eqs[2]{Eqs.~(\ref{#1}) and (\ref{#2})}
\newcommand\eqss[3]{Eqs.~(\ref{#1}), (\ref{#2}) and (\ref{#3})}

\def\to{\rightarrow}

\def\llgm{\left\lgroup}
\def\rrgm{\right\rgroup}
\def\lf{\left(}
\def\rg{\right)}
\newcommand\vev[1]{\langle {#1} \rangle}
\newcommand{\Gr}{\ensuremath{\widetilde{G}}}
\newcommand{\Yb}{\ensuremath{Y_{B}}}
\newcommand{\Yg}{\ensuremath{Y_{3/2}}}

\newcommand{\Vhi}{\ensuremath{\widehat V_{\rm HI}}}
\newcommand{\Hhi}{\ensuremath{\widehat H_{\rm HI}}}

\newcommand{\Khi}{\ensuremath{K}}

\newcommand{\mP}{\ensuremath{m_{\rm P}}}

\newcommand{\Mgut}{\ensuremath{M_{\rm GUT}}}
\newcommand{\Qef}{\ensuremath{\Lambda_{\rm UV}}}
\newcommand{\Ggut}{\ensuremath{G_{B-L}}}

\newcommand{\lm}{\ensuremath{\lambda_\mu}}

\def\openone{\leavevmode\hbox{\small1\kern-3.8pt\normalsize1}}
\newcommand{\dV}{\ensuremath{\Delta\widehat V_{\rm HI}}}
\newcommand{\Dex}{\ensuremath{\Delta_{\rm max\star}}}

\newcommand{\Gsn}{\ensuremath{\what{\Gamma}_{\rm \dph}}}
\newcommand{\GNsn}{\ensuremath{\what{\Gamma}_{\dph\to N_i^cN_i^c}}}
\newcommand{\Ghsn}{\ensuremath{\what{\Gamma}_{\dph\to \hu\hd}}}
\newcommand{\Gysn}{\ensuremath{\what{\Gamma}_{\dph\to XYZ}}}

\newcommand{\msn}{\ensuremath{\what m_{\rm \dph}}}

\newcommand{\hd}{{\ensuremath{H_d}}}
\newcommand{\hu}{{\ensuremath{H_u}}}

\newcommand{\dphi}{\ensuremath{\what{\delta\phi}}}
\newcommand{\dph}{\ensuremath{\delta\phi}}

\newcommand{\ks}{\ensuremath{k_\star}}
\newcommand{\Ns}{\ensuremath{{\what N_\star}}}
\newcommand{\ns}{\ensuremath{n_{\rm s}}}

\newcommand{\nb}{\ensuremath{N_{X}}}
\newcommand{\as}{\ensuremath{a_{\rm s}}}
\newcommand{\As}{\ensuremath{A_{\rm s}}}
\newcommand{\rw}{\ensuremath{r_{0.002}}}
\newcommand{\rs}{\ensuremath{r_{\pm}}}
\newcommand{\rpm}{\ensuremath{r_{\pm}}}

\newcommand{\rce}{\ensuremath{\widehat{\mathcal{R}}}}
\newcommand{\Ve}{\ensuremath{\widehat{V}}}

\newcommand{\sni}{\ensuremath{N^c_i}}
\newcommand{\ssni}{\ensuremath{\widetilde N_i^c}}

\newcommand{\aS}{\ensuremath{{\rm a}_S}}
\newcommand{\Ald}{\ensuremath{A_\lambda}}
\newcommand{\am}{\ensuremath{{\rm a}_{3/2}}}

\newcommand{\mrh[1]}{\ensuremath{M_{#1N^c}}}
\newcommand{\lrh[1]}{\ensuremath{\lambda_{#1N^c}}}

\newcommand{\mD[1]}{\ensuremath{m_{#1\rm D}}}
\newcommand{\mn[1]}{\ensuremath{m_{#1\nu}}}

\newcommand{\wrhn[1]}{\ensuremath{N^c_{#1}}}

\newcommand{\Whi}{\ensuremath{W_{\rm HI}}}

\newcommand{\Lg}{\ensuremath{\mathcal{L}}}

\def\ve{\varepsilon}

\def\bbet{{\bar\beta}}
\def\al{{\alpha}}
\def\bt{{\beta}}

\def\th{{\theta}}
\def\thb{{\bar\theta}}
\def\thn{{\theta_{\Phi}}}
\def\tb{{\tan\beta}}

\newcommand{\Trh}{\ensuremath{T_{\rm rh}}}
\newcommand{\sg}{\ensuremath{\phi}}

\newcommand{\ld}{\ensuremath{\lambda}}
\newcommand{\ldu}{\ensuremath{\uplambda}}
\newcommand{\Ld}{\ensuremath{\Lambda}}
\newcommand{\kp}{\ensuremath{\kappa}}

\newcommand{\sgx}{\ensuremath{\phi_\star}}
\newcommand{\sgf}{\ensuremath{\phi_{\rm f}}}

\newcommand{\what}{\ensuremath{\widehat}}
\newcommand{\wtilde}{\ensuremath{\widetilde}}

\newcommand{\se}{\ensuremath{\widehat \phi}}
\newcommand{\sex}{\ensuremath{\widehat{\phi}_\star}}

\newcommand{\geu}{\ensuremath{\widehat g}}
\newcommand{\eph}{\ensuremath{\widehat \epsilon}}

\newcommand{\mgr}{\ensuremath{m_{3/2}}}
\newcommand{\mg}{{\ensuremath{M_{1/2}}}}

\newcommand{\sign}{{\ensuremath{\rm sign}}}

\def\Kap{K\"{a}hler potential}

\def\Kaa{K\"{a}hler~}
\def\sub{subplanckian}
\def\sup{superpotential}
\def\str{Starobinsky}
\def\bcp{{\sc\small Bicep2}/{\it Keck Array}}
\newcommand{\plk}{{\it Planck}}

\newcommand{\diag}{\ensuremath{{\sf diag}}}
\newcommand{\im}{\ensuremath{{\sf Im}}}

\newcommand{\tr}{{\mbox{\sf\ssz T}}}

\newcommand{\cm}{\ensuremath{c_{-}}}
\newcommand{\cp}{\ensuremath{c_{+}}}
\newcommand{\fm}{\ensuremath{F_{-}}}
\newcommand{\fp}{\ensuremath{F_{+}}}

\newcommand{\nsu}{\ensuremath{{N_X}}}

\newcommand{\Gbl}{\ensuremath{G_{B-L}}}

\newcommand{\fr}{\ensuremath{f_{\cal R}}}
\newcommand{\frs}{\ensuremath{f_{n\star}}}

\newcommand{\fns}{\ensuremath{f_{n\star}}}

\newcommand{\fk}{\ensuremath{f_{\rm K}}}

\newcommand{\phc}{\ensuremath{\Phi}}
\newcommand{\phcb}{\ensuremath{\bar\Phi}}
\newcommand\mtta[4]{\mbox{
$\llgm\bem #1 &#2 \cr #3& #4\eem\rrgm$}}

\newcommand\mtn[9]{\ensuremath{\llgm\bem #1&#2&#3\cr #4&#5&#6 \cr #7&#8&#9\eem\rrgm}}

\newcommand{\bdhh}{{\ensuremath{\normalsize I{\kern-2.9pt H}}}}

\renewenvironment{subequations}{%
\refstepcounter{equation}%
\setcounter{parentequation}{\value{equation}}%
  \setcounter{equation}{0}
  \ignorespaces
}{%
  \setcounter{equation}{\value{parentequation}}%
  \ignorespacesafterend
}

\title{\LARGE\boldmath \bfseries\scshape Gravitational Waves, $\mu$ Term
\& Leptogenesis from $B-L$ Higgs Inflation in Supergravity}

\author{\Large \bfseries\scshape C. Pallis\\
Department of Physics, University of Cyprus, \\ P.O. Box 20537,
Nicosia 1678, CYPRUS\\ \vspace{3pt}
\email{cpallis@ucy.ac.cy}}

\abstract{We consider a renormalizable extension of the minimal
supersymmetric standard model endowed by an $R$ and a gauged $B -
L$ symmetry. The model incorporates chaotic inflation driven by a
quartic potential, associated with the Higgs field which leads to
a spontaneous breaking of $\bl$, and yields possibly detectable
gravitational waves. We employ quadratic \Kap s with a prominent
shift-symmetric part proportional to $\cm$ and a tiny violation,
proportional to $\cp$, included in a logarithm with prefactor
$-N<0$. An explanation of the $\mu$ term of the MSSM is also
provided, consistently with the low energy phenomenology, under
the condition that one related parameter in the superpotential is
somewhat small. Baryogenesis occurs via non-thermal leptogenesis
which is realized by the inflaton's decay to the lightest or
next-to-lightest right-handed neutrino with masses lower than
$1.8\cdot10^{13}~\GeV$. Our scenario can be confronted with the
current data on the inflationary observables, the baryon asymmetry
of the universe, the gravitino limit on the reheating temperature
and the data on the neutrino oscillation parameters, for
$0.012\lesssim\cp/\cm\lesssim1/N$ and gravitino as light as
$1~\TeV$.

\\ \\ {\ftn\sffamily {\scshape Keywords}:  Cosmology, Inflation, Supersymmetric Models} \\ {\ftn\sffamily {\scshape PACS codes}:  98.80.Cq,
12.60.Jv, 95.30.Cq, 95.30.Sf} \\\\ {\sl\bfseries Published in}
{\sl Universe} {\bf 4}, no. 1, 13 (2018)}

\begin{document}

\setcounter{page}{1} \pagestyle{fancyplain}

\addtolength{\headheight}{.5cm}

\rhead[\fancyplain{}{ \bf \thepage}]{\fancyplain{}{\sc
Gravitational Waves, $\mu$ Term \& nTL from $B-L$ HI in SUGRA}}
\lhead[\fancyplain{}{\sc \leftmark}]{\fancyplain{}{\bf \thepage}}
\cfoot{}

\section{Introduction}\label{intro}

One of the primary ideas, followed the introduction of inflation
\cite{guth} as a solution to longstanding cosmological problems --
such as the horizon, flatness and magnetic monopoles problems --,
was its connection with a phase transition related to the
breakdown of a \emph{Grand Unified Theory} ({\sf \ftn GUT}).
According to this economical and highly appealing scenario --
called henceforth \emph{Higgs inflation} ({\sf\ftn HI}) -- the
inflaton may be identified with one particle involved in the Higgs
sector \cite{old,jones2,nmH,susyhybrid,smHgut,moduliGUT} of a GUT
model. In a series of recent papers \cite{nMHkin, var} we
established a novel type of GUT-scale, mainly, HI called
\emph{kinetically modified non-Minimal HI}. This term is coined in
\cref{nMkin} due to the fact that, in the
non-\emph{Supersymmetric} ({\sf\ftn SUSY}) set-up, this
inflationary model, based on the $\phi^4$ power-law potential,
employs not only a suitably selected non-minimal coupling to
gravity $\fr=1+\cp\phi^{2}$ but also a kinetic mixing of the form
$\fk=\cm\fr^m$ -- cf. \cref{lee}. The merits of this construction
compared to the original (and certainly more predictive) model
\cite{old, nmi,atroest} of \emph{non-minimal inflation} ({\sf\ftn
nMI}) defined for $\fk=1$ are basically two:

\begin{itemize}

\item[{\sf (i)}] For $m\geq0$, the observables depend on the ratio
$r_\pm=\cp/\cm$ and can be done excellently consistent with the
the recent data \cite{plin, gwsnew} as regards the
tensor-to-scalar ratio, $r$. More specifically, all data taken by
the \bcp\ CMB polarization experiments up to and including the
2014 observing season ({\sf\ftn BK14}) \cite{gwsnew} seem to favor
$r$'s of order $0.01$, since the analysis yields
\beq r=0.028^{+0.026}_{-0.025}~~\Rightarrow~~0.003\lesssim
r\lesssim0.054 \hspace*{0.2cm} \mbox{at 68\%c.l.} \label{gws}\eeq

\item[{\sf (ii)}] The resulting theory respects the perturbative
unitarity \cite{cutoff, riotto} up to the Planck scale for
$r_{\pm}\leq1$ and $m$ of order unity.

\end{itemize}

In the SUSY -- which means \emph{Supergravity} ({\sf\ftn SUGRA})
-- framework the two ingredients necessary to achieve this kind of
nMI, i.e., the non-minimal kinetic mixing and coupling to gravity,
originate from the same function, the \Kap, and the set-up becomes
much more attractive. Actually, the non-minimal kinetic mixing and
gravitational coupling of the inflaton can be elegantly realized
introducing an approximate shift symmetry \cite{shift, shiftHI,
lee, jhep, nMHkin}. Namely, the constants $\cm$ and $\cp$
introduced above can be interpreted as the coefficients of the
principal shift-symmetric term ($\cm$) and its violation ($\cp$)
in the \Kap s $K$. Allowing also for a variation of the
coefficients of the logarithms appearing in the $K$'s we end up
with the most general form of these models analyzed in \cref{var}.

Here, we firstly single out the most promising models from those
investigated in \cref{var}, employing as a guiding principle the
consistency of the expansion of the $K$'s in powers of the various
fields. Namely, as we mention in \cref{var}, $m=0$ and $m=1$ are
the two most natural choices since they require just quadratic
terms in some of the $K$'s considered. From these two choices the
one with $m=1$ is privileged since it ensures $r$ within
\Eref{gws} with central value for the spectral index $\ns$. Armed
with the novel stabilization mechanisms for the non-inflaton
accompanied field -- recently proposed in the context of the
\str-type inflation \cite{su11} too --, we concentrate here on
$K$'s including \emph{exclusively} quadratic terms with $m=1$. The
embedding of the selected models in a complete framework is the
second aim of this paper. Indeed, a complete inflationary model
should specify the transition to the radiation domination, explain
the origin of the observed \emph{baryon asymmetry of the universe}
({\ftn\sf BAU}) \cite{baryo} and also, yield the \emph{minimal
supersymmetric standard model} ({\sf\ftn MSSM}) as low energy
theory. Although this task was carried out for similar models --
see, e.g., \crefs{nmH, R2r, quad} -- it would be certainly
interesting to try to adapt it to the present set-up. Further
restrictions are induced from this procedure.

A GUT based on $\Ggut=G_{\rm SM}\times U(1)_{B-L}$, where ${G_{\rm
SM}}= SU(3)_{\rm C}\times SU(2)_{\rm L}\times U(1)_{Y}$ is the
gauge group of the standard model and $B$ and $L$ denote the
baryon and lepton number respectively, consists \cite{jhep,
nMHkin, var} a conveniently simple framework which allows us to
exemplify our proposal. Actually, this is a minimal extension of
the MSSM which is obtained by promoting the already existing
$U(1)_{B-L}$ global symmetry to a local one. The Higgs fields
which cause the spontaneous breaking of the $\Ggut$ symmetry to
${G_{\rm SM}}$ can naturally play the role of inflaton. This
breaking provides large Majorana masses to the right-handed
neutrinos, $\sni$, whose the presence is imperative in order to
cancel the gauge anomalies and generate the tiny neutrino masses
via the seesaw mechanism. Furthermore, the out-of-equilibrium
decay of the $\sni$'s provides us with an explanation of the
observed BAU \cite{plcp} via \emph{non-thermal leptogenesis}
({\sf\ftn nTL}) \cite{lept} consistently with the gravitino
($\Gr$) constraint \cite{gravitino,brand,kohri,grNew} and the data
\cite{valle,lisi} on the neutrino oscillation parameters. As a
consequence, finally, of an adopted global $R$ symmetry, the
parameter $\mu$ appearing in the mixing term between the two
electroweak Higgs fields in the superpotential of MSSM is
explained as in \crefs{dvali, R2r} via the \emph{vacuum
expectation value} ({\sf\ftn v.e.v}) of the non-inflaton
accompanying field, provided that the relevant coupling constant
is rather suppressed.

Below, we present the particle content, the superpotential and the
possible \Kap s which define our model in Sec.~\ref{fhim}. In
Sec.~\ref{fhi} we describe the inflationary potential, derive the
inflationary observables and confront them with observations.
Sec.~\ref{secmu} is devoted to the resolution of the $\mu$ problem
of MSSM. In Sec.~\ref{pfhi} we analyze the scenario of nTL
exhibiting the relevant constraints and restricting further the
parameters. Our conclusions are summarized in Sec.~\ref{con}.
Throughout the text, the subscript of type $,z$ denotes derivation
\emph{with respect to} (w.r.t) the field $z$ and charge
conjugation is denoted by a star. Unless otherwise stated, we use
units where $\mP = 2.433\cdot 10^{18}~\GeV$ is taken unity.

\section{Model Description}\label{fhim}

We focus on an extension of MSSM invariant under the gauge group
$\Ggut$. Besides the MSSM particle content, the model is augmented
by six superfields: a gauge singlet $S$, three $\sni$'s, and a
pair of Higgs fields $\phc$ and $\phcb$ which break $\bl$. In
addition to the local symmetry, the model possesses also the
baryon and lepton number symmetries and a nonanomalous $R$
symmetry $U(1)_{R}$. The charge assignments under these symmetries
of the various matter and Higgs superfields are listed in
Table~\ref{tab1}. We below present the superpotential (\Sref{sup})
and (some of) the \Kap s (\Sref{ka}) which give rise to our
inflationary scenario.

\subsection{Superpotential} \label{sup}

The superpotential of our model naturally splits into two parts:
\beq W=W_{\rm MSSM}+\Whi,\>\>\>\mbox{where}\label{Wtot}\eeq
\paragraph{\sf\ftn (a)} $W_{\rm MSSM}$ is the part of $W$ which contains the
usual terms -- except for the $\mu$ term -- of MSSM, supplemented
by Yukawa interactions among the left-handed leptons ($L_i$) and $\sni$:
\beqs \beq W_{\rm MSSM} = h_{ijD} {d}^c_i {Q}_j \hd + h_{ijU}
{u}^c_i {Q}_j \hu+h_{ijE} {e}^c_i {L}_j \hd+ h_{ijN} \sni L_j \hu.
\label{wmssm}\eeq
Here the $i$th generation $SU(2)_{\rm L}$ doublet left-handed
quark and lepton superfields are denoted by $Q_i$ and $L_i$
respectively, whereas the $SU(2)_{\rm L}$ singlet antiquark
[antilepton] superfields by $u^c_i$ and ${d_i}^c$ [$e^c_i$ and
$\sni$] respectively. The electroweak Higgs superfields which
couple to the up [down] quark superfields are denoted by $\hu$
[$\hd$].

\paragraph{\sf\ftn (b)} $\Whi$ is the part of $W$ which is relevant for
HI, the generation of the $\mu$ term of MSSM and the Majorana
masses for $\sni$'s. It takes the form
\beq\label{Whi} \Whi= \ld S\lf \bar\Phi\Phi-M^2/4\rg+\lm
S\hu\hd+\lrh[i]\phcb N^{c2}_i\,. \eeq\eeqs
The imposed $U(1)_R$ symmetry ensures the linearity of $\Whi$
w.r.t $S$. This fact allows us to isolate easily via its
derivative the contribution of the inflaton into the F-term SUGRA
potential, placing $S$ at the origin -- see \Sref{fhi1}. It plays
also a key role in the resolution of the $\mu$ problem of MSSM via
the second term in the \emph{right-hand side} ({\sf\ftn r.h.s}) of
\Eref{Whi} -- see \Sref{secmu2}. The inflaton is contained in the
system $\bar\Phi - \Phi$. We are obliged to restrict ourselves to
\sub\ values of $\bar\Phi\Phi$ since the imposed symmetries do not
forbid non-renormalizable terms of the form $(\bar\Phi\Phi)^{p}$
with $p>1$ -- see \Sref{fhi2}. The third term in the r.h.s of
\Eref{Whi} provides the Majorana masses for the $\sni$'s -- cf.
\crefs{nmH, R2r, quad} -- and assures the decay \cite{Idecay} of
the inflaton to $\ssni$, whose subsequent decay can activate nTL.
Here, we work in the so-called \emph{$\sni$-basis}, where
$\mrh[i]$ is diagonal, real and positive. These masses, together
with the Dirac neutrino masses in Eq.~(\ref{wmssm}), lead to the
light neutrino masses via the seesaw mechanism.

\renewcommand{\arraystretch}{1.1}

\begin{table}[!t]
\begin{center}
\begin{tabular}{|c|c|c|c|c|}\hline
{\sc Superfields}&{\sc Representations}&\multicolumn{3}{|c|}{\sc
Global Symmetries}\\\cline{3-5}
&{\sc under $\Gbl$}& {\hspace*{0.3cm} $R$\hspace*{0.3cm} }
&{\hspace*{0.3cm}$B$\hspace*{0.3cm}}&{$L$} \\\hline\hline
\multicolumn{5}{|c|}{\sc Matter Fields}\\\hline
{$e^c_i$} &{$({\bf 1, 1}, 1, 1)$}& $1$&$0$ & $-1$ \\
{$N^c_i$} &{$({\bf 1, 1}, 0, 1)$}& $1$ &$0$ & $-1$
 \\
{$L_i$} & {$({\bf 1, 2}, -1/2, -1)$} &$1$&{$0$}&{$1$}
\\
{$u^c_i$} &{$({\bf 3, 1}, -2/3, -1/3)$}& $1$  &$-1/3$& $0$
\\
{$d^c_i$} &{$({\bf 3, 1}, 1/3, -1/3)$}& $1$ &$-1/3$& $0$
 \\
{$Q_i$} & {$({\bf \bar 3, 2}, 1/6 ,1/3)$} &$1$ &$1/3$&{$0$}
\\ \hline
\multicolumn{5}{|c|}{\sc Higgs Fields}\\\hline
{$\hd$}&$({\bf 1, 2}, -1/2, 0)$& {$0$}&{$0$}&{$0$}\\
{$\hu$} &{$({\bf 1, 2}, 1/2, 0)$}& {$0$} & {$0$}&{$0$}\\
\hline
{$S$} & {$({\bf 1, 1}, 0, 0)$}&$2$ &$0$&$0$  \\
{$\Phi$} &{$({\bf 1, 1}, 0, 2)$}&{$0$} & {$0$}&{$-2$}\\
{$\bar \Phi$}&$({\bf 1, 1}, 0,-2)$&{$0$}&{$0$}&{$2$}\\
\hline\end{tabular}
\end{center}
\caption[]{\sl \small The representations under $\Gbl$ and the
extra global charges of the superfields of our model.}\label{tab1}
\end{table}
\renewcommand{\arraystretch}{1.}

\subsection{\Kaa Potentials} \label{ka}

HI is feasible if $\Whi$ cooperates with \emph{one} of the following \Kap
s -- cf.~\cref{var}:
\beqs\bea
K_1&=&-N\ln\left(1+\cp\fp+F_{1X}(|X|^2)\right)+\cm\fm,\label{K1}\\
K_2&=&-N\ln\left(1+\cp\fp\right)+\cm\fm+F_{2X}(|X|^2),\label{K2}\\
K_3&=&-N\ln\left(1+\cp\fp\right)+F_{3X}(\fm, |X|^2), \label{K3}
\eea\eeqs
where $N>0$, $X^\gamma=S,\hu,\hd,\ssni$  and the complex scalar
components of the superfields $\Phi, \bar\Phi, S, \hu$ and $\hd$
are denoted by the same symbol whereas this of $\sni$ by $\ssni$.
The functions $F_\pm=\left|\Phi\pm\bar\Phi^*\right|^2$ assist us
in the introduction of shift symmetry for the Higgs fields  -- cf.
\cref{shiftHI, jhep}. In all $K$'s, $\fp$ is included in the
argument of a logarithm with coefficient $-N$ whereas $\fm$ is
outside it. As regards the non-inflaton fields $X^\gamma$, we
assume that they have identical kinetic terms expressed by the
functions $F_{lX}$ with $l=1,2,3$. In \Tref{tab0} we expose two
possible forms for each $F_{lX}$ following \cref{su11}. These are
selected so as to successfully stabilize the scalars $X^\gamma$ at
the origin employing only quadratic terms. Recall
\cite{linde1,su11} that the simplest term $|X|^2$ leads to
instabilities for $K=K_1$ and light excitations of $X^\gamma$ for
$K=K_2$ and $K_3$. The heaviness of these modes is required so
that the observed curvature perturbation is generated wholly by
our inflaton in accordance with the lack of any observational hint
\cite{plcp} for large non-Gaussianity in the cosmic microwave
background.

As we show in \Sref{fhi1}, the positivity of the kinetic energy of
the inflaton sector requires $\cp<\cm$ and $N>0$. For
$\rs=\cp/\cm\ll1$, our models are completely natural in the 't
Hooft sense because, in the limits $\cp\to0$ and $\ld\to0$, they
enjoy the following enhanced symmetries
\beq \Phi \to\ \Phi+c,\>\>\>\bar\Phi \to\ \bar\Phi+c^*
\>\>\>\mbox{and}\>\>\> X^\gamma \to\ e^{i\varphi_\gamma}
X^\gamma,\label{shift}\eeq
where $c$ and $\varphi_\gamma$ are complex and real numbers
respectively and no summation is applied over $\gamma$. This
enhanced symmetry has a string theoretical origin as shown in
\cref{lust}. In this framework, mainly integer $N$'s are
considered which can be reconciled with the observational data.
Namely, acceptable inflationary solutions are attained for $N=3$
[$N=2$] if $K=K_1$ [$K=K_2$ or $K_3$] -- see \Sref{fhi3}. However,
deviation of the $N$'s from these integer values is perfectly
acceptable \cite{jhep, var, roest, nIG} and can have a pronounced
impact on the inflationary predictions allowing for a covering of
the whole $\ns-r$ plane with quite natural values of the relevant
parameters.

\section{Inflationary Scenario}\label{fhi}

The salient features of our inflationary scenario are studied at
tree level in \Sref{fhi1} and at one-loop level in \Sref{fhi11}.
We then present its predictions in \Sref{fhi3}, calculating a
number of observable quantities introduced in Sec.~\ref{fhi2}.

\renewcommand{\arraystretch}{1.2}
\begin{table}[t]
\bec\begin{tabular}{|c|c|c|}\hline
$F_{lX}$&{\sc Exponential Form} & {\sc Logarithmic
Form}\\\hline\hline

$F_{1X}$ & $\exp\lf-|X|^2/N\rg-1$ & $-\ln(1+|X|^2/N)$ \\
$F_{2X}$ & $-N_X\lf \exp\lf-|X|^2/N_X\rg-1\rg$ & $N_X\ln(1+|X|^2/N_X)$ \\
$F_{3X}$ & $-N_X\lf \exp\lf-\cm\fm/N_X-|X|^2/N_X\rg-1\rg$& $N_X\ln(1+\cm\fm/N_X+|X|^2/N_X)$ \\
\hline
\end{tabular}\eec
\caption{\sl \small Functional forms of $F_{lX}$ with $l=1,2,3$
shown in the definition of $K_1, K_2$ and $K_3$ in
\eqss{K1}{K2}{K3} respectively.\label{tab0}}
\end{table}
\renewcommand{\arraystretch}{1}

\subsection{Inflationary Potential}\label{fhi1}

Within SUGRA the \emph{Einstein frame} ({\sf\small EF}) action for
the scalar fields $z^\al=S,\phc,\phcb,\hu,\hd$ and $\wtilde N_i^c$
can be written as
\beqs \beq\label{Saction1} {\sf S}=\int d^4x \sqrt{-\what{
\mathfrak{g}}}\lf-\frac{1}{2}\rce +K_{\al\bbet}\geu^{\mu\nu} D_\mu
z^\al D_\nu z^{*\bbet}-\Ve\rg, \eeq
where $\rce$ is the Ricci scalar and $\mathfrak{g}$ is the
determinant of the background Friedmann-Robertson-Walker metric,
$g^{\mu\nu}$ with signature $(+,-,-,-)$. We adopt also the
following notation
\beq \label{KDm} K_{\al\bbet}={\Khi_{,z^\al
z^{*\bbet}}}>0\>\>\>\mbox{and}\>\>\>D_\mu z^\al=\partial_\mu
z^\al+ig A^{\rm a}_\mu T^{\rm a}_{\al\bt} z^\bt\eeq
are the covariant derivatives for the scalar fields $z^\al$. Also,
$g$ is the unified gauge coupling constant, $A^{\rm a}_\mu$ are
the vector gauge fields and $T^{\rm a}$ are the generators of the
gauge transformations of $z^\al$. Also $\Ve$ is the EF SUGRA
scalar potential which can be found via the formula
\beq \Ve=\Ve_{\rm F}+ \Ve_{\rm D}\>\>\>\mbox{with}\>\>\> \Ve_{\rm
F}=e^{\Khi}\left(K^{\al\bbet}D_\al \Whi D^*_\bbet \Whi^*-3{\vert
\Whi\vert^2}\right) \>\>\>\mbox{and}\>\>\>\Ve_{\rm D}=
{1\over2}g^2 \sum_{\rm a} {\rm D}_{\rm a} {\rm D}_{\rm
a},\label{Vsugra} \eeq
where we use the notation
\beq \label{Kinv} K^{\al\bbet}K_{\al\bar
\gamma}=\delta^\bbet_{\bar \gamma},\>\>D_\al W_{\rm HI} =W_{{\rm
HI},z^\al} +K_{\al}W_{\rm HI}~~\mbox{and}~~{\rm D}_{\rm
a}=z^\al\lf T_{\rm a}\rg_\al^\bt K_\bt~~\mbox{with}~~
K_{\al}={\Khi_{,z^\al}}.\eeq\eeqs

If we express $\Phi, \bar\Phi$ and $X^\gamma= S,\hu,\hd,\ssni$
according to the parametrization
\beq\label{hpar} \Phi=\frac{\sg e^{i\th}}{\sqrt{2}}
\cos\thn,~~\bar\Phi=\frac{\sg e^{i\thb}}{\sqrt{2}}
\sin\thn~~\mbox{and}~~X^\gamma= \frac{x^\gamma +i\bar
x^\gamma}{\sqrt{2}}\,,\eeq
where $0\leq\thn\leq\pi/2$, we can easily deduce from
\Eref{Vsugra} that a D-flat direction occurs at
\beq \label{inftr} x^\gamma=\bar
x^\gamma=\th=\thb=0\>\>\>\mbox{and}\>\>\>\thn={\pi/4}\eeq
along which the only surviving term in \Eref{Vsugra} can be
written universally as
\beq \label{Vhi} \Vhi= e^{K}K^{SS^*}\, |W_{{\rm
HI},S}|^2=\frac{\ld^2(\sg^2-M^2)^2}{16\fr^{2(1+n)}}\>\>\>\mbox{where}\>\>\>\fr=1+\cp\sg^2\eeq
plays the role of a non-minimal coupling to Ricci scalar in the
\emph{Jordan frame} ({\sf\ftn JF}) -- see \crefs{linde1, jhep}.
Also, we set \beq \label{ndef} n= \left\{\bem
%
(N-3)/2\hfill\cr
N/2-1\hfill\cr \eem
\right. \>\>\>\mbox{and}\>\>\> K^{SS^*}=\left\{\bem
%
\fr\hfill\cr
1\hfill\cr \eem
\right. \>\>\>\mbox{for}\>\>\> \left\{\bem
%
K=K_1\,,\hfill\cr
K=K_{2} ~~\mbox{and}~~ K_3\,.\hfill\cr \eem
\right.\eeq

The introduction of $n$ allows us to obtain a unique inflationary
potential for all the $K$'s in Eqs.~(\ref{K1}) -- (\ref{K3}). For
$K=K_1$ and $N=3$ or $K=K_2$ or $K_3$ and $N=2$ we get $n=0$ and
$\Vhi$ develops an inflationary plateau as in the original case of
non-minimal inflation \cite{old}. Contrary to that case, though,
here we have also $\cm$ which dominates the canonical
normalization of $\phi$ -- see \Sref{fhi11} -- and allows for
distinctively different inflationary outputs as shown in
\crefs{nMkin, nMHkin}. Finally, the variation of $n$ above and
below zero allows for more drastic deviations \cite{jhep, var}
from the predictions of the original model \cite{old}. In
particular, for $n<0$, $\Vhi$ remains increasing function of
$\phi$, whereas for $n>0$, $\Vhi$ develops a local maximum
\beq \Vhi(\sg_{\rm max})=\frac{\ld^2 n^{2 n}}{16 \cp^2(1 + n)^{2
(1 + n)}}~~\mbox{at}~~ \sg_{\rm max}=\frac1{\sqrt{\cp
n}}\,.\label{Vmax}\eeq
In a such case we are forced to assume that hilltop \cite{lofti}
HI occurs with $\sg$ rolling from the region of the maximum down
to smaller values. Therefore, a mild tuning of the initial
conditions is required which can be quantified somehow defining
\cite{gpp} the quantity
\beq \Dex=\left(\sg_{\rm max} - \sgx\right)/\sg_{\rm
max}\,,\label{dms}\eeq
$\sgx$ is the value of $\sg$ when the pivot scale $\ks=0.05/{\rm
Mpc}$ crosses outside the inflationary horizon. The naturalness of
the attainment of HI increases with $\Dex$ and it is maximized
when $\sg_{\rm max}\gg\sgx$ which result to $\Dex\simeq1$.

\begin{figure}[!t]\vspace*{-.25in}
\begin{center}
\epsfig{file=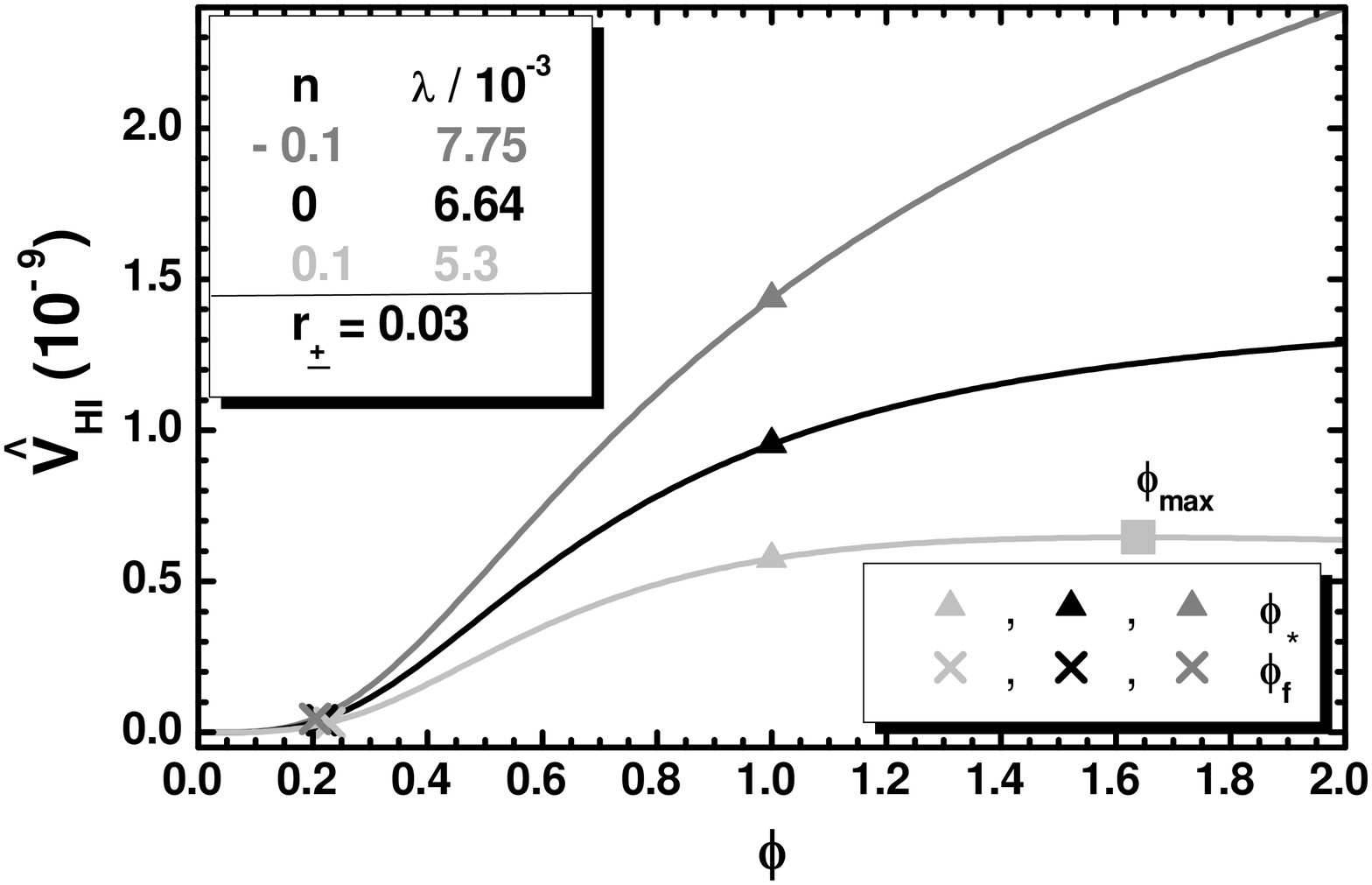,height=3.65in,angle=-90}\eec
\vspace*{-.2in}\hfill  \caption[]{\sl \small The inflationary
potential $\Vhi$ as a function of $\sg$ for $\sg>0$, and
$\rs\simeq0.03$, and $n=-0.1$, $\ld=7.75\cdot 10^{-3}$ (gray
line), $n=0$, $\ld=6.64\cdot 10^{-3}$ (black line), or $n=+0.1$,
$\ld=5.3\cdot 10^{-3}$ (light gray line). The values of $\sgx$,
$\sgf$ and $\sg_{\rm max}$ (for $n=1/10$) are also
indicated.}\label{fig2}
\end{figure}

The structure of $\Vhi$ as a function of $\sg$ is displayed in
\Fref{fig2}. We take $\sgx=1$, $\rs=0.03$ and $n=-0.1$ (light gray
line), $n=0$ (black line) and $n=0.1$ (gray line). Imposing the
inflationary requirements mentioned in \Sref{fhi3} we find the
corresponding values of $\ld$ and $\cm$ which are $(7.75,
6.64~{\rm or}~5.3)\cdot 10^{-3}$ and $(1.7, 1.46,{\rm
or}~1.24)\cdot 10^2$ respectively. The corresponding observable
quantities are found numerically to be $\ns=0.971, 0.969$ or
$0.966$ and $r=0.045, 0.03$ or $0.018$ with
$\as\simeq-5\cdot10^{-4}$ in all cases. We see that $\Vhi$ is a
monotonically increasing function of $\sg$  for $n\leq0$ whereas
it develops a maximum at $\sg_{\rm max}=1.64$, for $n=0.1$, which
leads to a mild tuning of the initial conditions of HI since
$\Dex=39\%$, according to the criterion introduced above. It is
also remarkable that $r$ increases with the inflationary scale,
$\Vhi^{1/4}$, which, in all cases, approaches the SUSY GUT scale
$M_{\rm GUT}\simeq 8.2\cdot10^{-3}$ facilitating the
interpretation of the inflaton as a GUT-scale Higgs field.

\subsection{Stability and one-Loop Radiative Corrections}\label{fhi11}

As deduced from \Eref{Vhi} $\Vhi$ is independent from $\cm$ which
dominates, though, the canonical normalization of the inflaton. To
specify it together with the normalization of the other fields, we
note that, for all $K$'s in Eqs.~(\ref{K1}) -- (\ref{K3}),
$K_{\al\bbet}$ along the configuration in \Eref{inftr} takes the
form
\beqs\beq \label{Kab} \lf K_{\al\bbet}\rg=\diag\lf
M_\pm,\underbrace{K_{\gamma\bar\gamma},...,K_{\gamma\bar\gamma}}_{8~\mbox{\ftn
elements}}\rg\eeq with \beq
M_\pm=\frac{1}{\fr^2}\mtta{\kappa}{\bar\kappa}{\bar\kappa}{\kappa}
~~\mbox{and}~~K_{\gamma\bar\gamma}=\left\{\bem
%
1/\fr\hfill\cr
1\hfill\cr \eem
\right. \>\>\>\mbox{for}\>\>\> \left\{\bem
%
K=K_1\,,\hfill\cr
K=K_{2} ~~\mbox{and}~~ K_3\,.\hfill\cr \eem
\right.\label{Sni1} \eeq\eeqs
Here $\kp=\cm\fr^2-N\cp$ and $\bar\kp={N\cp^2\sg^2}$. Upon
diagonalization of $M_\pm$ we find its eigenvalues which are
\beq
\label{kpm}\kp_+=\cm\lf1+N\rpm(\cp\sg^2-1)/{\fr^2}\rg\simeq\cm
\>\>\>\mbox{and}\>\>\> \kp_-=\cm\lf1- {N\rpm}/{\fr}\rg ,\eeq
where the positivity of $\kp_-$ is assured during and after HI for
\beq \label{rsmin}\rs<\fr/N~~\mbox{with}~~\rs=\cp/\cm\,.\eeq
Given that $\fr>1$ and $\vev{\fr}\simeq1$, \Eref{rsmin} implies
that the maximal possible $\rs$ is $\rs^{\rm max}\simeq1/N$. Given
that $N$ tends to $3$ [$2$] for $K=K_1$ [$K=K_2$ or $K_3$], the
inequality above discriminates somehow the allowed parameter space
for the various choices of $K$'s in Eqs.~(\ref{K1}) -- (\ref{K2}).

Inserting \eqs{hpar}{Sni1} in the second term of the r.h.s of
\Eref{Saction1} we can, then, specify the EF canonically
normalized fields, which are denoted by hat, as follows
\beqs\bea \nonumber K_{\al\bbet}\dot z^\al \dot z^{*\bbet}&=&
{\kp_+\over 2}\lf\dot \sg^2+{1\over2}\sg^2\dot\theta^2_+
\rg+{\kp_-\over 2}\sg^2\lf{1\over2}\dot\theta^2_-
+\dot\theta^2_\Phi \rg+\frac12K_{\gamma\bar\gamma}\lf\dot
x^{\gamma2}+\dot{\bar
x}^{\gamma2}\rg\\&\simeq&\frac12\lf\dot{\widehat
\sg}^{~2}+\dot{\widehat \th}_+^{~2}+\dot{\widehat
{\th}}_-^{~2}+\dot{\widehat
\th}_\Phi^{~2}+\dot{\what{x}}^{\gamma2}+\dot{\what{\bar{x}}}^{~~\gamma
2}\rg, \label{Snik}\eea
where $\th_{\pm}=\lf\bar\th\pm\th\rg/\sqrt{2}$ and the dot denotes
derivation w.r.t the cosmic time $t$. The hatted fields of the
$\phc-\phcb$ system can be expressed in terms of the initial
(unhatted) ones via the relations
\beq \label{VJe}
\frac{d\se}{d\sg}=J=\sqrt{\kp_+},\>\>\widehat{\theta}_+
={J\over\sqrt{2}}\sg\theta_+,\>\>\widehat{\theta}_-
=\sqrt{\frac{\kp_-}{2}}\sg\theta_-,\>\>\mbox{and}\>\>\widehat
\theta_\Phi =
\sqrt{\kp_-}\sg\lf\theta_\Phi-\frac{\pi}{4}\rg\,\cdot\eeq
As regards the non-inflaton fields, the (approximate)
normalization is implemented as follows
\beq (\what{x}^{\gamma},\what{\bar
x}^{\gamma})=\sqrt{K_{\gamma\bar\gamma}}(x^\gamma,\bar
x^\gamma).\eeq\eeqs
As we show below, the masses of the scalars besides $\se$ during
HI are heavy enough such that the dependence of the hatted fields
on $\sg$ does not influence their dynamics -- see also \cref{nmH}.

We can verify that the inflationary direction in \Eref{inftr} is
stable w.r.t the fluctuations of the non-inflaton fields. To this
end, we construct the mass-squared spectrum of the scalars taking
into account the canonical normalization of the various fields in
\Eref{Snik} -- for details see \cref{jhep}. In the limit
$\cm\gg\cp$, we find the expressions of the masses squared $\what
m^2_{z^\al}$ (with $z^\al=\theta_+,\theta_\Phi,x^\gamma$ and $\bar
x^\gamma$) arranged in \Tref{tab3}. These results approach rather
well the quite lengthy, exact expressions taken into account in
our numerical computation.  The various unspecified there
eigenvalues are defined as follows
\beqs\beq h_\pm=(h_u\pm{h_d})/\sqrt{2},\>\>\> {\bar h}_\pm=({\bar
h}_u\pm{\bar h}_d)/\sqrt{2}\>\>\>\mbox{and}\>\>\>\what \psi_\pm
=(\what{\psi}_{\Phi+}\pm \what{\psi}_{S})/\sqrt{2}, \eeq
where the (unhatted) spinors $\psi_\Phi$ and $\psi_{\bar\Phi}$
associated with the superfields $\Phi$ and $\bar\Phi$ are related
to the normalized (hatted) ones in \Tref{tab3} as follows
\beq \label{psis}
\what\psi_{\Phi\pm}=\sqrt{\kp_\pm}\psi_{\Phi\pm}\>\>\>
\mbox{with}\>\>\>\psi_{\Phi\pm}=(\psi_\Phi\pm\psi_{\bar\Phi})/\sqrt{2}\,.
\eeq\eeqs

From \Tref{tab3} it is evident that $0<\nsu\leq6$ assists us to
achieve $m^2_{{s}}>\Hhi^2=\Vhi/3$ -- in accordance with the
results of \cref{su11} -- and also enhances the ratios
$m^2_{X^{\tilde\gamma}}/\Hhi^2$ for
$X^{\tilde\gamma}=\hu,\hd,\ssni$ w.r.t the values that we would
have obtained, if we had used just canonical terms in the $K$'s.
On the other hand, $\what m^2_{h-}>0$ requires
\beqs\bea \label{lm1} &\lm<\ld(1+\cp\sg^2/N)/4\lf1/\sg^2+\cp\rg
&~~\mbox{for}~~K=K_1;\\
\label{lm2} & \lm<\ld\sg^2(1+1/\nsu)/4&
~~\mbox{for}~~K=K_2~~\mbox{and}~~K_3\,.\eea\eeqs
In both cases, the quantity in the r.h.s of the inequality takes
its minimal value at $\sg=\sgf$ and numerically equals to
$2\cdot10^{-5}-10^{-6}$. Similar numbers are obtained in
\cref{R2r} although that higher order terms in the \Kap\ are
invoked there. We do not consider such a condition on $\lm$ as
unnatural, given that $h_{1U}$ in \Eref{wmssm} is of the same
order of magnitude too -- cf. \cref{fermionM}. Note that the due
hierarchy in \eqs{lm1}{lm2} between $\lm$ and $\ld$ differs from
that imposed in the models \cite{dvali} of F-term hybrid
inflation, where $S$ plays the role of inflaton and
$\phc,\phcb,~\hu$ and $\hd$ are confined at zero. Indeed, in that
case we demand \cite{dvali} $\lm>\ld$ so that the tachyonic
instability in the $\phc-\phcb$ direction occurs first, and the
$\phc-\phcb$ system start evolving towards its v.e.v, whereas
$\hu$ and $\hd$ continue to be confined to zero. In our case,
though, the inflaton is included in the $\bar\Phi-\Phi$ system
while $S$ and the $\hu-\hd$ system are safely stabilized at the
origin both during and after HI. Therefore, $\phi$ is led at its
vacuum whereas $S$, $\hu$ and $\hd$ take their non-vanishing
electroweak scale v.e.vs afterwards.

\renewcommand{\arraystretch}{1.4}
\begin{sidewaystable}[h!]
\bec\begin{tabular}{|c|c|c|c|c|c|}\hline
{\sc Fields}&{\sc Eigen-} & \multicolumn{4}{c|}{\sc Masses
Squared}\\\cline{3-6}
&{\sc states}&& {$K=K_1$}&{$K=K_2$}&{$K=K_{3}$} \\
\hline\hline
14 Real &$\widehat\theta_{+}$&$\widehat m_{\theta+}^2$&
\multicolumn{2}{|c|}{$6\Hhi^2$} &$6(1+1/\nsu)\Hhi^2$\\\cline{3-6}
Scalars&$\widehat \theta_\Phi$ &$\widehat m_{ \theta_\Phi}^2$&
\multicolumn{2}{|c|}{$M^2_{BL}+6\Hhi^2$}&
$M^2_{BL}+6(1+1/\nsu)\Hhi^2$\\\cline{3-6}
&$\widehat s, \widehat{\bar{s}}$ & $\widehat m_{
s}^2$&$6\cp\sg^2\Hhi^2/N$&\multicolumn{2}{c|}{$6\Hhi^2/\nsu$}\\\cline{3-6}
& $\widehat h_{\pm},\widehat{\bar h}_{\pm}$ &  $ \widehat
m_{h\pm}^2$&$3\Hhi^2\lf1+{\cp\sg^2}/{N}\pm{4\lm}(1/{\sg^2}+\cp)/{\ld}\rg$&\multicolumn{2}{c|}{$3\Hhi^2\lf1+1/\nsu\pm{4\lm}/{\ld\sg^2}\rg$}\\\cline{3-6}
& $\widehat{\tilde\nu}^c_{i}, \widehat{\bar{\tilde\nu}}^c_{i}$ &
$\widehat m_{i\tilde
\nu^c}^2$&$3\Hhi^2\lf1+{\cp\sg^2}/{N}+{16\ld^2_{iN^c}}(1/{\sg^2}+\cp)/{\ld^2}\rg$&\multicolumn{2}{c|}{$3\Hhi^2\lf1+1/\nsu+16\ld^2_{iN^c
}/\ld^2\sg^2\rg$}\\\cline{2-6}
1 Gauge Boson & $A_{BL}$ &  $
M_{BL}^2$&\multicolumn{3}{c|}{$g^2\cm\lf1-N\rpm
/\fr\rg\sg^2$}\\\hline
$7$ Weyl & $\what \psi_\pm $ & $\what m^2_{ \psi\pm}$ &
$6\lf(N-3)\cp\sg^2-2\rg^2\Hhi^2/\cm\sg^2\fr^{2}$&\multicolumn{2}{c|}{$6\lf(N-2)\cp\sg^2-2\rg^2\Hhi^2/\cm\sg^2\fr^{2}$}\\\cline{3-6}
Spinors &${N_i^c}$& $ \widehat
m_{{iN^c}}^2$&\multicolumn{3}{c|}{$48\ld^2_{iN^c
}\Hhi^2/\ld^2\sg^2$}\\\cline{2-6} &$\ldu_{BL},
\widehat\psi_{\Phi-}$&
$M_{BL}^2$&\multicolumn{3}{c|}{$g^2\cm\lf1-N\rpm /\fr\rg\sg^2$}\\
\hline
\end{tabular}\eec
\hfill \caption{\sl\small The mass squared spectrum of our models
along the inflationary trajectory in \Eref{inftr} for
$K=K_1,K_2,K_3$ and $\sg\ll1$. To avoid very lengthy formulas, we
neglect terms proportional to $M\ll\sg$.}\label{tab3}
\renewcommand{\arraystretch}{1.}
\end{sidewaystable}
\clearpage

In \Tref{tab3} we display also the mass $M_{BL}$ of the gauge
boson which becomes massive having `eaten'  the Goldstone boson
$\th_-$. This signals the fact that $\Ggut$ is broken during HI.
Shown are also the masses of the corresponding fermions -- note
that the fermions $\tilde{h}_\pm$ and $\tilde{\bar{h}}_\pm$,
associated with $h_\pm$ and $\bar{h}_\pm$ remain massless. The
derived mass spectrum can be employed in order to find the
one-loop radiative corrections, $\dV$ to $\Vhi$. Considering SUGRA
as an effective theory with cutoff scale equal to $\mP$, the
well-known Coleman-Weinberg formula \cite{cw} can be employed
self-consistently taking into account the masses which lie well
below $\mP$, i.e., all the masses arranged in \Tref{tab3} besides
$M_{BL}$ and $\what m_{\th_\Phi}$. Therefore, the one-loop
correction to $\Vhi$ reads
\bea\dV&=&{1\over64\pi^2}\lf \widehat m_{\th+}^4\ln{\widehat
m_{\th+}^2\over\Lambda^2} +2 \widehat m_{s}^4\ln{\widehat
m_{s}^2\over\Lambda^2} + 4\widehat m_{h+}^4\ln{\widehat
m_{h+}^2\over\Lambda^2}+ 4 \widehat m_{h-}^4\ln{\widehat
m_{h-}^2\over\Lambda^2}
\nonumber \right.\\
&+&\left.2\sum_{i=1}^3\lf \widehat m_{i\tilde \nu^c}^4\ln{\widehat
m_{i\tilde \nu^c}^2\over\Lambda^2}-\widehat m_{
iN^c}^4\ln{\widehat m_{iN^c}^2\over\Lambda^2}\rg-4\widehat
m_{\psi_{\pm}}^4\ln{\widehat m_{\psi_{\pm}}^2\over\Lambda^2}\rg
,\label{Vhic}\eea
where $\Lambda$ is a \emph{renormalization group} ({\sf\ftn RG})
mass scale. The resulting $\dV$ lets intact our inflationary
outputs, provided that $\Lambda$ is determined by requiring
$\dV(\sgx)=0$ or $\dV(\sgf)=0$. These conditions yield
$\Ld\simeq3.2\cdot10^{-5}-1.4\cdot10^{-4}$ and render our results
practically independent of $\Lambda$ since these can be derived
exclusively by using $\Vhi$ in \Eref{Vhi} with the various
quantities evaluated at $\Ld$ -- cf. \cref{jhep}. Note that their
renormalization-group running is expected to be negligible because
$\Ld$ is close to the inflationary scale
$\Vhi^{1/4}\simeq(3-7)\cdot10^{-3}$ -- see \Fref{fig2}.

\subsection{Inflationary Observables}\label{fhi2}

A period of slow-roll HI is determined by the condition -- see
e.g. \cref{review}
\beq{\ftn\sf
max}\{\widehat\epsilon(\sg),|\widehat\eta(\sg)|\}\leq1,\label{sr}\eeq
where\beqs\beq\label{sr1}\widehat\epsilon=
{1\over2}\left(\frac{\Ve_{{\rm
HI},\se}}{\Vhi}\right)^2={1\over2J^2}\left(\frac{\Ve_{{\rm
HI},\sg}} {\Vhi}\right)^2\simeq \frac{8(1-n\cp\sg^2)^2}{\cm\sg^2
\fr^{2}}\eeq
and \beq \widehat\eta = \frac{\Ve_{{\rm HI},\se\se}}{\Vhi}
={1\over J^2}\left( \frac{\Ve_{{\rm
HI},\sg\sg}}{\Vhi}-\frac{\Ve_{{\rm HI},\sg}}{\Vhi}{J_{,\sg}\over
J}\right)=4\:\frac{3 - 3(1+3n)\cp\sg^2 +n(1+ 4
n)\cp^2\sg^4}{\cm\sg^2 \fr^{2}}\,\cdot \label{sr2}\eeq\eeqs
Expanding $\widehat\epsilon$ and $\widehat\eta$ for $\sg\ll 1$ we
can find from \Eref{sr} that HI terminates for $\sg=\sgf$ such
that
\beq \sgf\simeq\mbox{\ftn\sf max}\left\{\frac{2
\sqrt{2/\cm}}{\sqrt{1+ 16(1 + n)\rs}},\frac{2
\sqrt{3/\cm}}{\sqrt{1+ 36 (1+n)\rs}}\right\}. \label{sgap}\eeq

The number of e-foldings, $\Ns$, that the pivot scale
$\ks=0.05/{\rm Mpc}$ suffers during HI can be calculated through
the relation
\begin{equation}
\label{Nhi}  \Ns=\:\int_{\se_{\rm f}}^{\se_\star}\, d\se\:
\frac{\Ve_{\rm HI}}{\Ve_{\rm HI,\se}}\simeq\begin{cases}
({(1+\cp\sgx^2)^{2}-1})/{16\rs}~~&\mbox{for}~~n=0\,,
\\ -\lf{n \cp \sgx^2 +
(1 + n)\ln(1 - n \cp\sgx^2)}\rg/{8 n^2
\rs}~~&\mbox{for}~~n\neq0\,,
\end{cases}
\end{equation}
where $\sex$ is the value of $\se$ when $\ks$ crosses the
inflationary horizon. As regards the consistency of the relation
above for $n>0$, we note that we get $n \cp\sgx^2<1$ in all
relevant cases and so, $\ln(1 - n \cp\sgx^2)<0$ assures the
positivity of $\Ns$. Given that $\sgf\ll\sgx$, we can write $\sgx$
as a function of $\Ns$ as follows
\begin{equation}
\label{sgxb}\sgx\simeq
{\sqrt{\frs-1\over\cp}}~~\mbox{with}~~\frs=\begin{cases}
\lf1+16\rs\Ns\rg^{1/2}~~&\mbox{for}~~n=0\,,\\
\lf(1+n)/n\rg\lf1+W_k\lf{y/(1+n)}\rg\rg~~&\mbox{for}~~n\neq0\,.
\end{cases}
\end{equation}
Here $W_k$ is the Lambert $W$ or product logarithmic function
\cite{wolfram} and the parameter $y$ is defined as
$y=-\exp\lf-(1+8n^2\Ns\rs)/(1+n)\rg$. We take $k=0$ for $n\geq0$
and $k=-1$ for $n<0$. We can impose a lower bound on $\cm$ above
which $\sgx\leq1$ for every $\rs$. Indeed, from \Eref{sgxb} we
have
\begin{equation}
\label{cmmin}\sgx\leq1~~\Rightarrow~~\cm\geq{\lf\fns-1\rg}/{\rs}
\end{equation}
and so, our proposal can be stabilized against corrections from
higher order terms of the form $(\phc\phcb)^p$ with $p>1$ in
$\Whi$ -- see \Eref{Whi}. Despite the fact that $\cm$ may take
relatively large values, the corresponding effective theory is
valid up to $\mP=1$ -- contrary to the pure quartic nMI
\cite{cutoff, riotto}. To clarify further this point we have to
identify the ultraviolet cut-off scale $\Qef$ of theory analyzing
the small-field behavior of our models.  More specifically, we
expand about $\vev{\phi}=M\ll1$ the second term in the r.h.s of
\Eref{Saction1} for $\mu=\nu=0$ and $\Vhi$ in \Eref{Vhi}. Our
results can be written in terms of $\se$ as
\beqs\bea\label{Jexp}  J^2
\dot\phi^2&\simeq&\lf1+3N\rs^2\what{\sg}^2-5N\rs^3\what{\sg}^4+\cdots\rg\dot\se^2;\\
\Vhi&\simeq&\frac{\ld^2\what{\sg}^4}{16\cm^{2}}\lf1-2(1+n)\rs\what{\sg}^{2}+(3+5n)\rs^2\what{\sg}^4-\cdots\rg\,.
\label{Vexp}\eea\eeqs
From the expressions above we conclude that $\Qef=\mP$ since
$\rs\leq1$ due to \Eref{rsmin}. Although the expansions presented
above, are valid only during reheating we consider the extracted
$\Qef$ as the overall cut-off scale of the theory since the
reheating is regarded \cite{riotto} as an unavoidable stage of HI.

The power spectrum $\As$ of the curvature perturbations generated
by $\sg$ at the pivot scale $\ks$ is estimated as follows
\beq \label{Proba}\sqrt{\As}=\: \frac{1}{2\sqrt{3}\, \pi} \;
\frac{\Ve_{\rm HI}(\sex)^{3/2}}{|\Ve_{\rm
HI,\se}(\sex)|}\simeq\frac{
\ld\sqrt{\cm}}{32\sqrt{3}\pi}\frac{\sgx^3 \fr(\sgx)^{-n}}{1- n\cp
\sgx^2}\,\cdot \eeq
The resulting relation reveals that $\ld$ is proportional to $\cm$
for fixed $n$ and $\rs$. Indeed, plugging \Eref{sgxb} into the
expression above, we find
\beq \ld=32\sqrt{3\As}\pi
\cm\rs^{3/2}\frs^n\frac{n(1-\frs)+1}{(\frs-1)^{3/2}}\,\cdot
\label{lan}\eeq

At the same pivot scale, we can also calculate $\ns$, its running,
$\as$, and $r$ via the relations
\beqs\baq \label{ns} && \ns=\: 1-6\widehat\epsilon_\star\ +\
2\widehat\eta_\star\simeq 1-4n^2\rs-2n\frac{\rs^{1/2}}{\Ns^{1/2}}-\frac{3-2n}{2\Ns}-\frac{3-n}{8(\Ns^3\rs)^{1/2}}\,, \>\>\> \\
&& \label{as} \as =\:{2\over3}\left(4\widehat\eta_\star^2-(n_{\rm
s}-1)^2\right)-2\widehat\xi_\star\simeq-\frac{n\rs^{1/2}}{\Ns^{3/2}}-\frac{3-2n}{2\Ns^2},\\
&& \label{rt} r=16\widehat\epsilon_\star\simeq
-\frac{8n}{\Ns}+\frac{3+2n}{6\Ns^2\rs}+\frac{6-n}{3(\Ns^3\rs)^{1/2}}
+\frac{8n^2\rs^{1/2}}{\Ns^{1/2}}\,, \eaq\eeqs
where $\widehat\xi={\Ve_{\rm HI,\widehat\sg} \Ve_{\rm
HI,\widehat\sg\widehat\sg\widehat\sg}/\Ve_{\rm HI}^2}$ and the
variables with subscript $\star$ are evaluated at $\sg=\sgx$.

\subsection{Comparison with Observations}\label{fhi3}

The approximate analytic expressions above can be verified by the
numerical analysis of our model. Namely, we apply the accurate expressions in
\eqs{Nhi}{Proba} and confront the corresponding observables with the requirements
\cite{plcp}
\begin{equation}
\label{Ntot} \mbox{\ftn\sf
(a)}\>\>\Ns\simeq61.5+\ln{\Vhi(\sgx)^{1/2}\over\Vhi(\sgf)^{1/4}}+\frac12\fr(\sgx)~~~\mbox{and}~~~\mbox{\ftn\sf
(b)}\>\>\As^{1/2}\simeq4.627\cdot10^{-5}\,.\eeq
We, thus, restrict $\ld$ and $\sgx$ and compute the model
predictions via \eqss{ns}{as}{rt} for any selected $n$ and $\rs$.
In \sEref{Ntot}{a} we consider an equation-of-state parameter
$w_{\rm int}=1/3$ correspoding to quartic potential which is
expected to approximate rather well $\Vhi$ for $\sg\ll1$. For
rigorous comparison with observations we compute
$\rw=16\eph(\se_{0.002})$ where $\se_{0.002}$ is the value of
$\se$ when the scale $k=0.002/{\rm Mpc}$, which undergoes $\what
N_{0.002}=\Ns+3.22$ e-foldings during HI, crosses the horizon of
HI. These must be in agreement with the fitting of the \plk,
\emph{Baryon Acoustic Oscillations} ({\sf\ftn BAO}) and \bcp\ data
\cite{plin,gwsnew} with $\Lambda$CDM$+r$ model, i.e.,
\begin{equation}  \label{nswmap}
\mbox{\ftn\sf
(a)}\>\>\ns=0.968\pm0.009\>\>\>~\mbox{and}\>\>\>~\mbox{\ftn\sf
(b)}\>\>r\leq0.07,
\end{equation}
at 95$\%$ \emph{confidence level} ({\sf\ftn c.l.}) with
$|\as|\ll0.01$.

\begin{figure}[!t]\vspace*{-.12in}
\hspace*{-.19in}
\begin{minipage}{8in}
\epsfig{file=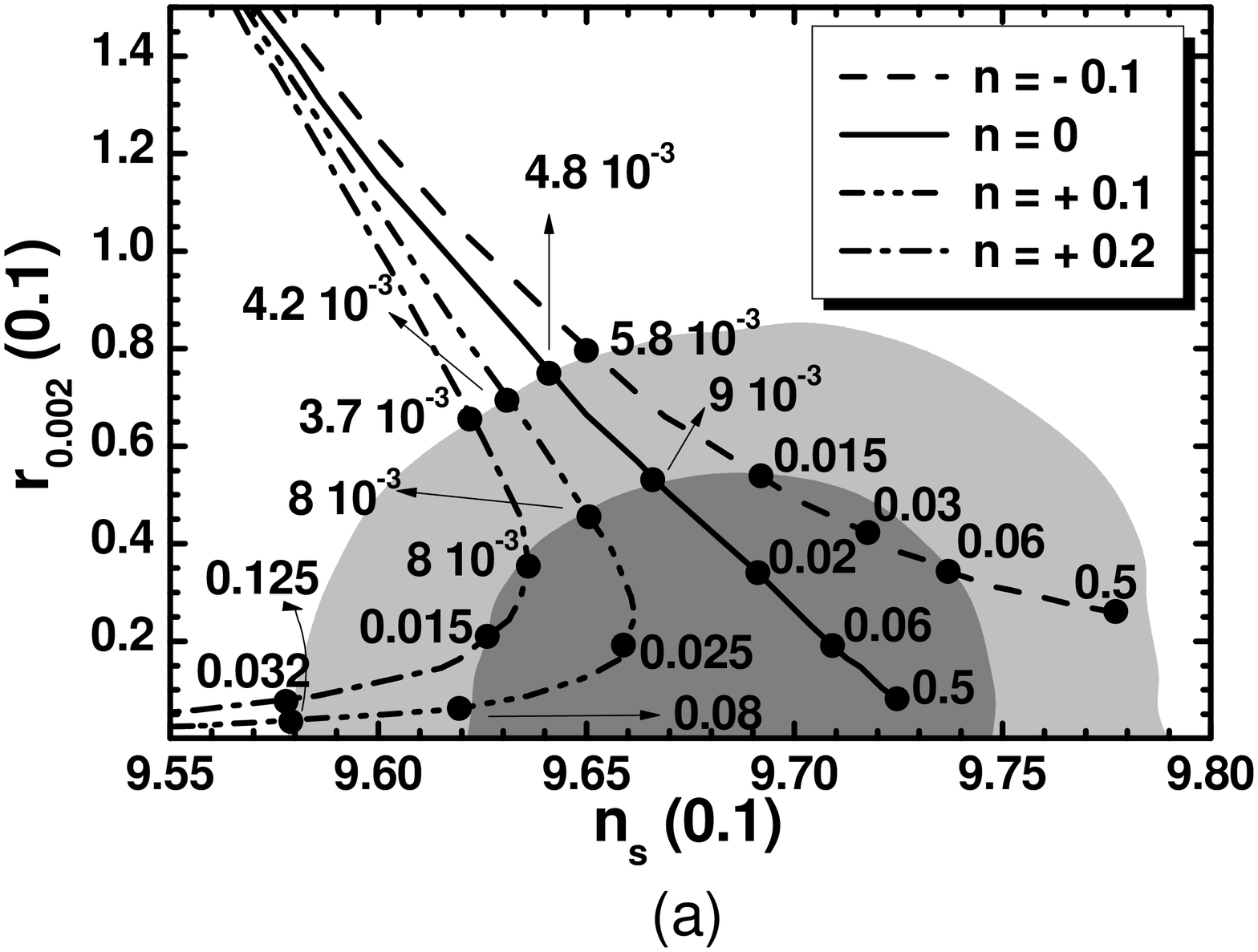,height=3.6in,angle=-90}
\hspace*{-1.2cm}
\epsfig{file=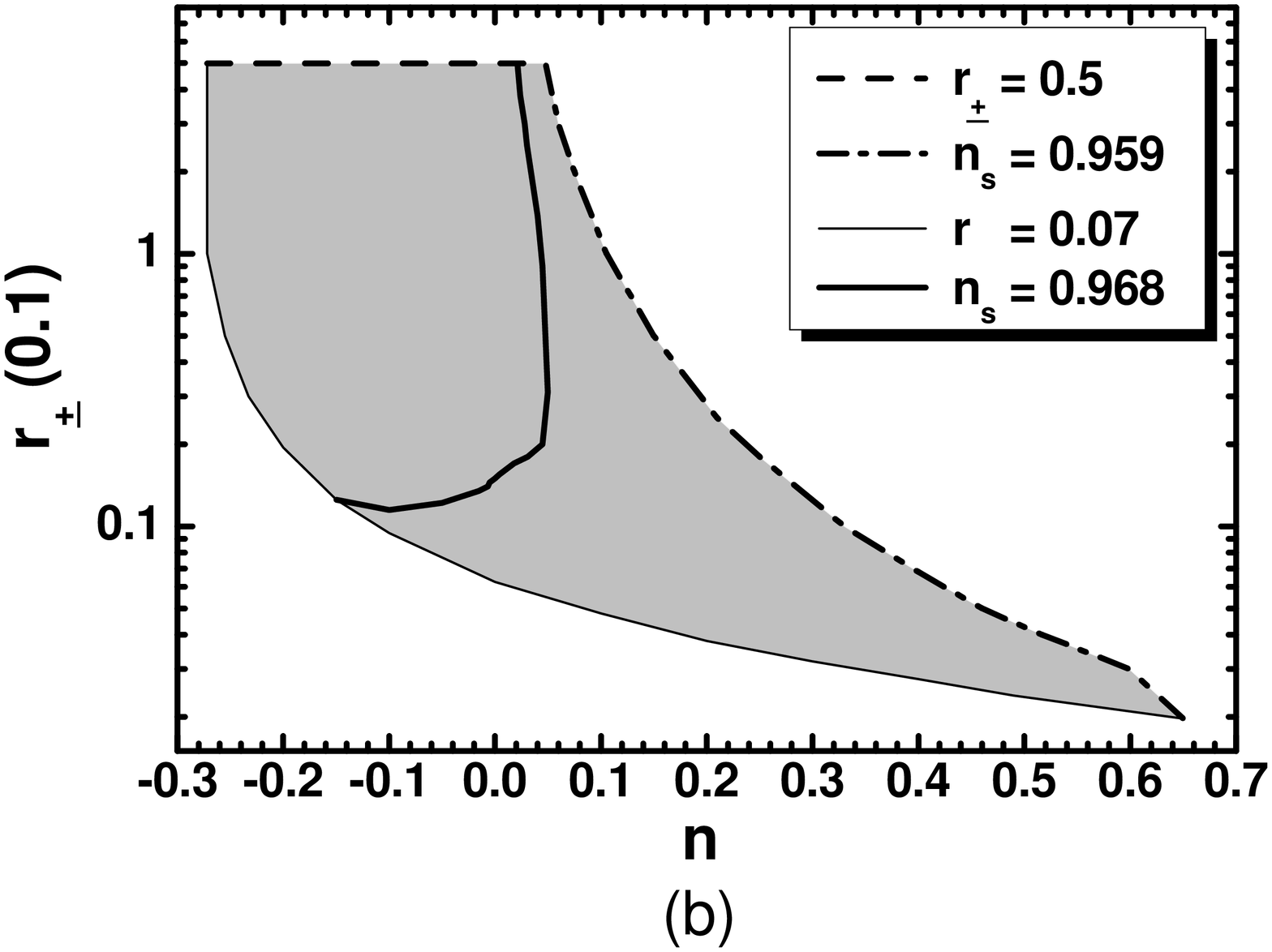,height=3.6in,angle=-90} \hfill
\end{minipage}
\hfill \caption{\sl\small {\sffamily\ftn (a)} Allowed curves in
the $\ns-\rw$ plane for $K=K_2$ and $K_3$, $n=-0.1, 0, 0.1, 0.2$
with the $\rs$ values indicated on the curves -- the marginalized
joint $68\%$ [$95\%$] regions from \plk, BAO and BK14 data are
depicted by the dark [light] shaded contours. {\sffamily\ftn (b)}
Allowed (shaded) regions in the $n-\rs$ plane for $K=K_2$ and
$K_3$. The conventions adopted for the various lines are
shown.}\label{fig1}
\end{figure}\renewcommand{\arraystretch}{1.}


Let us clarify here that the free parameters of our models are
$n$, $\rs$ and $\ld/\cm$ and not $n$, $\cm$, $\cp$ and $\ld$ as
naively expected. Indeed, if we perform the rescalings
\beq\label{resc}
\Phi\to\Phi/\sqrt{\cm},~~\bar\Phi\to\bar\Phi/\sqrt{\cm}~~~\mbox{and}~~~
S\to S,\eeq
$W$ in \Eref{Whi} depends on $\ld/\cm$ and the $K$'s in \Eref{K1} --
(\ref{K3}) depend on $n$ and $\rs$. As a consequence, $\Vhi$
depends exclusively on $\ld/\cm$, $n$ and $\rs$. Since the
$\ld/\cm$ variation is rather trivial -- see \cref{nMHkin} -- we
focus on the variation of the other parameters.

Our results are displayed in \Fref{fig1}. Namely, in
\sFref{fig1}{a} we show a comparison of the models' predictions
against the observational data \cite{plin,gwsnew} in the $\ns-\rw$
plane. We depict the theoretically allowed values with dot-dashed,
double dot-dashed, solid and dashed lines for $n=0.2,0.1,0$ and
$-0.1$ respectively. The variation of $\rs$ is shown along each
line. For low enough $\rs$'s -- i.e. $\rs\leq0.0005$ -- the
various lines converge to $(\ns,\rw)\simeq(0.947,0.28)$ obtained
within \emph{minimal} quartic inflation defined for $\cp=0$.
Increasing $\rs$ the various lines enter the observationally
allowed regions, for $\rs$ equal to a minimal value $\rs^{\rm
min}$, and cover them. The lines corresponding to $n=0,-0.1$
terminate for $\rs=\rs^{\rm max}\simeq0.5$, beyond which
\Eref{rsmin} is violated. Finally, the lines drawn with $n=0.2$ or
$n=0.1$ cross outside the allowed corridors and so the
$r_{\pm}^{\rm max}$'s, are found at the intersection points. From
\sFref{fig1}{a} we infer that the lines with $n>0$ [$n<0$] cover
the left lower [right upper] corner of the allowed range. As we
anticipated in \Sref{fhi1}, for $n>0$ HI is of hilltop type. The
relevant parameter $\Dex$ ranges from $0.07$ to $0.66$ for $n=0.1$
and from $0.19$ to $0.54$ for $n=0.2$ where $\Dex$ increases as
$\rs$ drops. That is, the required tuning is not severe mainly for
$\rs<0.1$.

As deduced from \sFref{fig1}{a}, the observationally favored
region can be wholly filled varying conveniently $n$ and $\rs$. It
would, therefore, interesting to delineate the allowed region of
our models in the $n-\rs$ plane, as shown in \sFref{fig1}{b}. The
conventions adopted for the various lines are also shown in the
legend of the plot. In particular, the allowed (shaded) region is
bounded by the dashed line, which originates from \Eref{rsmin},
and the dot-dashed and thin lines along which the lower and upper
bounds on $\ns$ and $r$ in \Eref{nswmap} are saturated
respectively. We remark that increasing $\rs$ with $n=0$, $r$
decreases, in accordance with our findings in \sFref{fig1}{a}. On
the other hand, $\rs$ takes more natural -- in the sense of the
discussion at the end of \Sref{ka} -- values (lower than unity)
for larger values of $|n|$ where hilltop HI is activated. Fixing
$\ns$ to its central value in \sEref{nswmap}{a} we obtain the
thick solid line along which we get clear predictions for
$(n,\rs)$ and so, the remaining inflationary observables. Namely,
we find
\beq\label{res1} -1.21\lesssim {n\over0.1}\lesssim0.215,\>\>\>
0.12\lesssim {\rs\over0.1}\lesssim5,\>\>\> 0.4\lesssim
{r\over0.01}\lesssim7\>\>\>\mbox{and}\>\>\>0.25\lesssim
10^5{\ld\over \cm}\lesssim2.6\,. \eeq Hilltop HI is attained for
$0<n\leq0.0215$ and there, we get $\Dex\gtrsim0.4$. The parameter
$\as$ is confined in the range $-(5-6)\cdot10^{-4}$ and so, our
models are consistent with the fitting of data with the
$\Lambda$CDM+$r$ model \cite{plin}. Moreover, our models are
testable by the forthcoming experiments like {\sc Bicep3}
\cite{bcp3}, PRISM \cite{prism} and LiteBIRD \cite{bird} searching
for primordial gravity waves since $r\gtrsim0.0019$.

Had we employed $K=K_1$, the various lines ended at $\rs\simeq0.5$
in \sFref{fig1}{a} and the allowed region in \sFref{fig1}{b} would
have been shortened until $\rs\simeq0.33$. This bound would have
yielded slightly larger $\rw^{\rm min}$'s. Namely, $\rw^{\rm
min}\simeq0.0084$ or $0.026$ for $n=0$ or $-0.1$ respectively --
the $\rw^{\rm min}$'s for $n>0$ are let unaffected. The lower
bound of $r/0.01$ and the upper ones on $\rs/0.1$ and
$10^5\ld/\cm$ in \Eref{res1} become $0.64$, $3.3$ and $2.1$
whereas the bounds on $\as$ remain unaltered.

\section{Higgs Inflation and $\mu$ Term of MSSM}
\label{secmu}

A byproduct of the $R$ symmetry associated with our model is that
it assists us to understand the origin of $\mu$ term of MSSM. To
see how this works, we first -- in \Sref{secmu1} -- derive the
SUSY potential of our models, and then  -- in \Sref{secmu2} --  we
study the generation of the $\mu$ parameter and investigate the
possible consequences for the phenomenology of MSSM  -- see
\Sref{pheno}. Here and henceforth we restore units, i.e., we take
$\mP=2.433\cdot10^{18}~\GeV$.

\subsection{SUSY Potential}\label{secmu1}

Since $\Vhi$ in \Eref{Vhi} is non-renormalizable, its SUSY limit
$V_{\rm SUSY}$ depends not only on $\Whi$ in \Eref{Whi}, but also
on the $K$'s in Eqs.~(\ref{K1}) -- (\ref{K3}). In particular,
$V_{\rm SUSY}$ turns out to be \cite{martin}
\beqs \beq \label{Vsusy} V_{\rm SUSY}= \widetilde K^{\al\bbet}
W_{\rm HI\al} W^*_{\rm HI\bbet}+\frac{g^2}2 \mbox{$\sum_{\rm a}$}
{\rm D}_{\rm a} {\rm D}_{\rm a}\,,\eeq
where $\widetilde K$ is the limit of the aforementioned $K$'s for $\mP\to\infty$ which is
\beq \label{Kquad}\widetilde K=\cm F_- -N\cp F_+
+|S|^2+|\hu|^2+|\hd|^2 +|\ssni|^2\,.\eeq
Upon substitution of $\widetilde K$ into \Eref{Vsusy} we obtain
\bea \nonumber && V_{\rm
SUSY}=\ld^2\left|\phcb\phc-\frac14{M^2}+\frac{\lm}{\ld}\hu\hd\right|^2
+\frac{1}{\cm(1-N\rs)}\lf\left|\ld S\phc+\ld_{iN^c}\widetilde
N^{c2}\right|^2+\ld^2|S\phcb|^2\rg+\\&& \lm^2
|S|^2\lf|\hu|^2+|\hd|^2\rg+4\ld_{iN^c}^2|\phcb\ssni|^2
+\frac{g^2}2\lf\cm(1-N\rs)\lf|\phc|^2-|\phcb|^2\rg+|\ssni|^2\rg^2\,.
\label{VF}\eea\eeqs  From the last equation, we find that the SUSY
vacuum lies along the D-flat direction $|\phcb|=|\phc|$ with
\beq \vev{S}=\vev{\hu}=\vev{\hd}=\vev{\ssni}=0
\>\>\>\mbox{and}\>\>\>
|\vev{\Phi}|=|\vev{\bar\Phi}|=M/2\,.\label{vevs} \eeq
As a consequence, $\vev{\Phi}$ and $\vev{\bar\Phi}$ break
spontaneously $U(1)_{B-L}$ down to $\mathbb{Z}^{B-L}_2$. Since
$U(1)_{B-L}$ is already broken during HI, no cosmic string are
formed -- contrary to what happens in the models of the standard
F-term hybrid inflation \cite{susyhybrid, dvali, gpp}, which
employ $\Whi$ in \Eref{Whi} too.

\subsection{Generation of the $\mu$ Term of MSSM}\label{secmu2}

The contributions from the soft SUSY breaking terms, although
negligible during HI, since these are much smaller than $\sg$, may
shift \cite{dvali,R2r} slightly $\vev{S}$ from zero in
\Eref{vevs}. Indeed, the relevant potential terms are
\beq V_{\rm soft}= \lf\ld A_\ld S \phcb\phc+\lm A_\mu S \hu\hd +
\ld_{iN^c} A_{iN^c}\phc \widetilde N^{c2}_i- {\rm a}_{S}S\ld M^2/4
+ {\rm h. c.}\rg+ m_{\gamma}^2\left|X^\gamma\right|^2,
\label{Vsoft} \eeq
where $m_{\gamma}, A_\ld, A_\mu, A_{iN^c}$ and $\aS$ are soft SUSY
breaking mass parameters.  Rotating $S$ in the real axis by an
appropriate $R$-transformation, choosing conveniently the phases
of $\Ald$ and $\aS$ so as the total low energy potential $V_{\rm
tot}=V_{\rm SUSY}+V_{\rm soft}$ to be minimized -- see \Eref{VF}
-- and substituting in $V_{\rm soft}$ the SUSY v.e.vs of $\phc,
\phcb, \hu, \hd$ and $\sni$ from \Eref{vevs} we get
\beqs\beq \vev{V_{\rm tot}(S)}= \ld^2\,M^2S^2/2\cm(1-N\rs)-\ld\am
\mgr M^2 S, \label{Vol} \eeq
where we take into account that $m_S\ll M$ and we set $|A_\ld| +
|{\rm a}_{S}|=2\am\mgr$ with $\mgr$ being the $\Gr$ mass and
$\am>0$ a parameter of order unity which parameterizes our
ignorance for the dependence of $|A_\ld|$ and $|{\rm a}_{S}|$ on
$\mgr$.  The minimization condition for the total potential in
\Eref{Vol} w.r.t $S$ leads to a non vanishing $\vev{S}$ as follows
\beq \label{vevS}\frac{d}{d S} \vev{V_{\rm
tot}(S)}=0~~\Rightarrow~~\vev{S}\simeq \am\mgr\cm(1-N\rs)/\ld.\eeq
At this $S$ value, $\vev{V_{\rm tot}(S)}$ develops a minimum since
\beq \label{Vss}\frac{d^2}{d S^2} \vev{V_{\rm
tot}(S)}=\ld^2\,M^2/\cm(1-N\rs)\eeq\eeqs
becomes positive for $\rs<1/N$, as dictated by \Eref{rsmin}. Let
us emphasize here that SUSY breaking effects explicitly break
$U(1)_R$ to the $\mathbb{Z}_2^{R}$ matter parity, under which all
the matter (quark and lepton) superfields change sign. Combining
$\mathbb{Z}_2^{R}$ with the $\mathbb{Z}_2^{\rm f}$ fermion parity,
under which all fermions change sign, yields the well-known
$R$-parity. Recall that this residual symmetry prevents the rapid
proton decay, guarantees the stability of the \emph{lightest SUSY
particle} ({\ftn\sf LSP}) and therefore, it provides a
well-motivated \emph{cold dark matter} ({\ftn\sf CDM})  candidate.
Since $S$ has the $R$ symmetry of $W$, $\vev{S}$ in \Eref{vevS}
breaks also spontaneously $U(1)_R$ to $\mathbb{Z}_2^{R}$. Thanks
to this fact, $\mathbb{Z}_2^{R}$ remains unbroken and so, no
disastrous domain walls are formed.

The generated $\mu$ term from the second term in the r.h.s of
\Eref{Whi} is \beqs\beq\mu =\lm \vev{S}
\simeq\lm\am\mgr\cm(1-N\rs)/\ld\label{mu}\eeq
which, taking into account \Eref{lan}, is written as
\beq\mu\simeq1.2\cdot10^{2}\lm\am\mgr(1-N\rs)
\rs^{-3/2}\frs^{-n}\frac{(\frs-1)^{3/2}}{n(1-\frs)+1}\,,\label{mu1}\eeq\eeqs
where $\cm$ and $\ld$ are eliminated. As a consequence, the
resulting $\mu$ in \Eref{mu} depends on $\rs$ and $n$ but does not
depend on $\ld$ and $\cm$ -- in contrast to the originally
proposed scheme in \cref{dvali} where a $\ld$ dependence remains.
Note, also, that $\lm$ (and so $\mu$) may have either sign without
any essential alteration in the stability analysis of the
inflationary system -- see \Tref{tab3}. Thanks to the magnitude of
the proportionality constant and given that $\rs^{-3/2}$ turns out
to be about $10^3$ for $\rs$ of order $0.01$, as indicated by
\Fref{fig1}, we conclude that any $|\mu|$ value is accessible for
the $\lm$ values allowed by \eqs{lm1}{lm2} without any ugly
hierarchy between $\mgr$ and $\mu$.

To highlight further the statement above, we can employ \Eref{mu}
to derive the $\mgr$ values required so as to obtain a specific
$\mu$ value. E.g., we fix $\mu=1~\TeV$ as suggested by many MSSM
versions for acceptable low energy phenomenology -- see
\cref{mssm}. Given that \Eref{mu} dependents on $\rs$ and $n$,
which crucially influences $n$ and $r$, we expect that the
required $\mgr$ is a function of $n$ and $r$ as depicted in
\sFref{fig3}{a} and \sFref{fig3}{b} respectively. We take
$\lm=10^{-6}$, in accordance with \eqs{lm1}{lm2}, $\am=1$, $K=K_2$
or $K_3$ with $\nsu=2$ and $n=-0.1$ (dot-dashed line), $n=0$
(solid line), or $n=+0.1$ (dashed line). Varying $\rs$ in the
allowed ranges indicated in \sFref{fig1}{a} for any of the $n$'s
above we obtain the variation of $\mgr$ solving \Eref{mu} w.r.t
$\mgr$. We see that $\mgr\geq1.6~\TeV$ with the lowest value
obtained for $n=0.1$.  Also, $\mgr$ corresponding to $n=0$ and
$-0.1$ increases sharply as $\rs$ approaches $0.49$ due to the
denominator $1-N\rs$ which approaches zero. Had we used $K=K_1$
this enhancement would have been occurred as $\rs$ tends to
$0.33$.

Obviously the proposed resolution of the $\mu$ problem of MSSM
relies on the existence of non-zero $\Ald$ and/or $\aS$. These
issues depend on the adopted model of SUSY breaking. Here we have
in mind mainly the gravity mediated SUSY breaking without, though,
to specify the extra terms in the \sup\ and the \Kap s which
ensure the appropriate soft SUSY breaking parameters and the
successful stabilization of the sgolstino -- cf.~\cref{buch}.
Since this aim goes beyond the framework of this work, we restrict
ourselves to assume that these terms can be added without
disturbing the inflationary dynamics.

\subsection{Connection with the MSSM Phenomenology}\label{pheno}

\begin{figure}[!t]\vspace*{-.12in}
\hspace*{-.19in}
\begin{minipage}{8in}
\epsfig{file=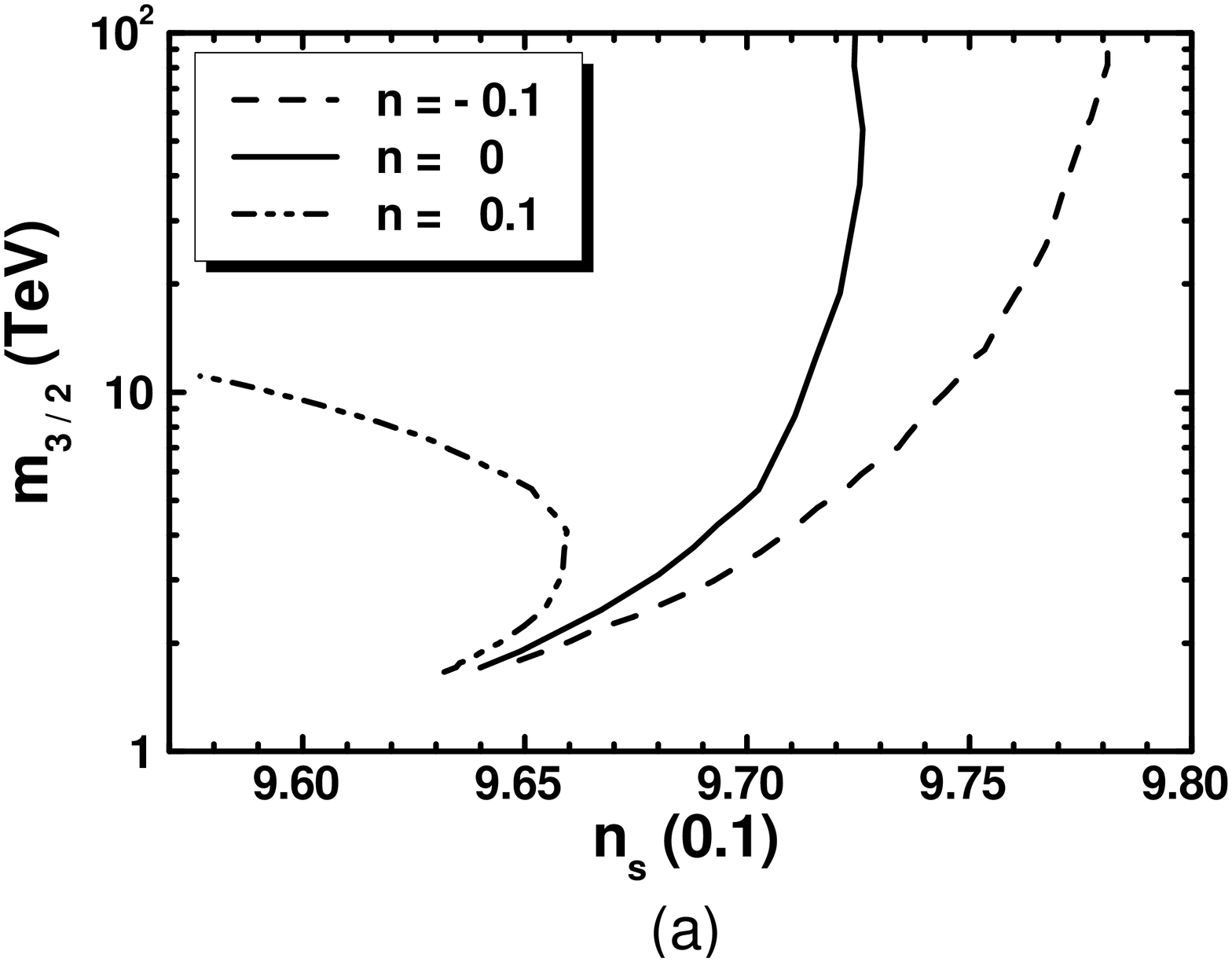,height=3.6in,angle=-90}
\hspace*{-1.2cm}
\epsfig{file=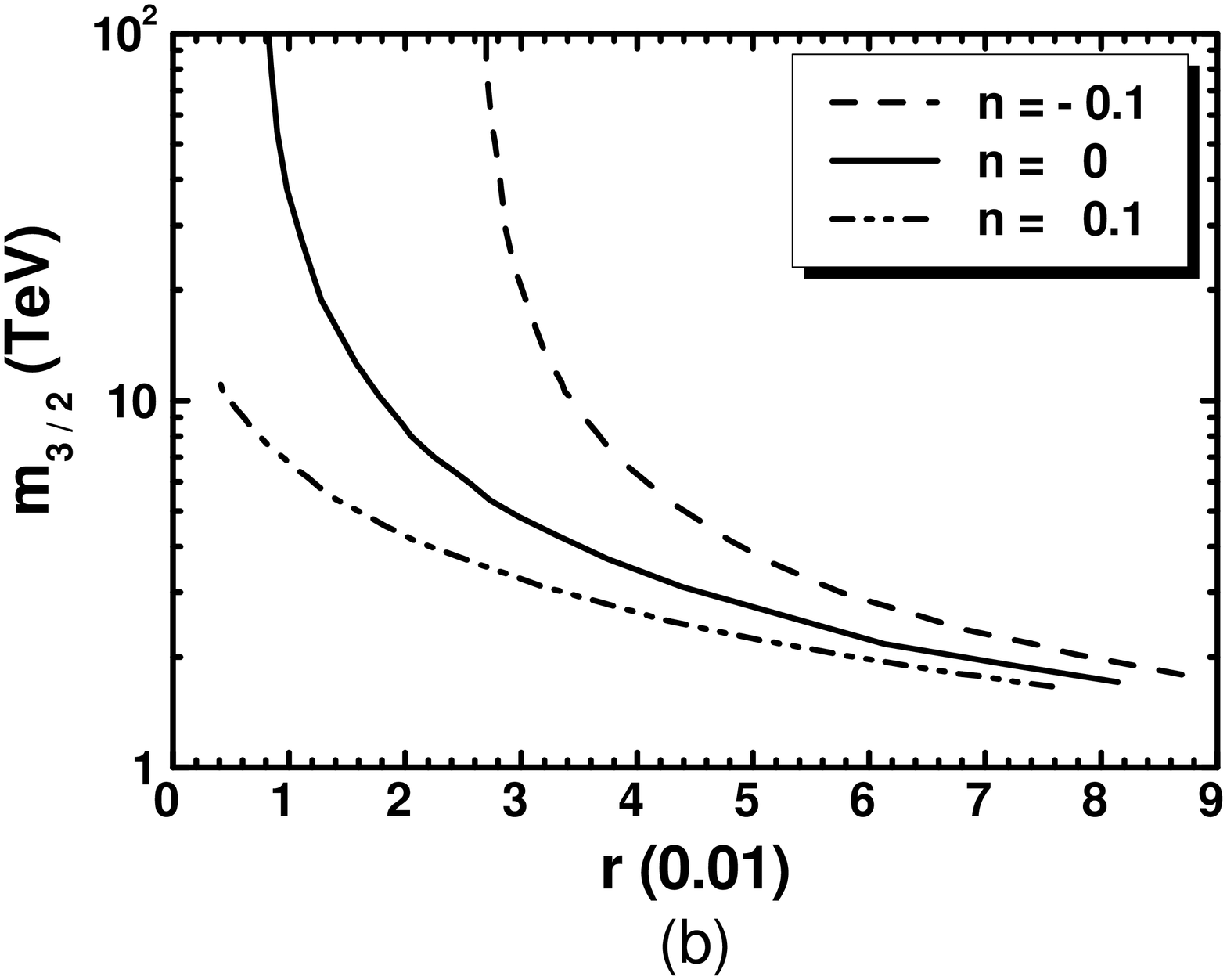,height=3.6in,angle=-90} \hfill
\end{minipage}
\hfill \caption{\sl\small The gravitino mass $\mgr$ versus $\ns$
{\sffamily\ftn (a)} or $r$ {\sffamily\ftn (b)} for $\mu=1~\TeV,
\lm=10^{-6}$, $\am=1$,  $K=K_2$ or $K_3$ with $\nsu=2$ and
$n=-0.1$ (dot-dashed line), $n=0$ (solid line), or $n=+0.1$
(dashed line).}\label{fig3}
\end{figure}

Taking advantage from the updated investigation of the parameter
space of \emph{Constrained MSSM} ({\ftn\sf CMSSM}) in \cref{mssm}
we can easily verify that the $\mu$ and $\mgr$ values satisfying
\Eref{mu} are consistent with the values required by the analyses
of the low energy observables of MSSM. We concentrate on CMSSM
which is the most predictive, restrictive and well-motivated
version of MSSM, employing the free parameters
$$\sign\mu,~~\tan\beta=\vev{\hu}/\vev{\hd},~~\mg,~~m_0~~\mbox{and}~~A_0,$$
where $\sign\mu$ is the sign of $\mu$, and the three last mass
parameters denote the common gaugino mass, scalar mass, and
trilinear coupling constant, respectively, defined at a high scale
which is determined by the unification of the gauge coupling
constants. The parameter $|\mu|$ is not free, since it is computed
at low scale enforcing the conditions for the electroweak symmetry
breaking. The values of these parameters can be tightly restricted
imposing a number of cosmo-phenomenological constraints. Namely,
these constraints originate from the cold dark matter abundance in
the universe and its direct detection experiments, the
$B$-physics, as well as the masses of the sparticles and the
lightest neutral CP-even Higgs boson. Some updated results are
recently presented in \cref{mssm}, where we can also find the
best-fit values of $|A_0|$, $m_0$ and $|\mu|$ listed in
\Tref{tab}.  We see that there are four allowed regions
characterized by the specific mechanism for suppressing the relic
density of the lightest sparticle which can act as dark matter. If
we identify $m_0$ with $\mgr$ and $|A_0|$ with $|A_\ld|=|\aS|$ we
can derive first $\am$ and then the $\lm$ values which yield the
phenomenologically desired $|\mu|$. Here we assume that
renormalization effects in the derivation of $\mu$ are negligible.
For the completion of this calculation we have to fix some sample
values of $(n,\rs)$. From those shown in \Eref{res1}, we focus on
this which is favored from the String theory with $n=0$ and this
which assure central values of the observables in
\eqs{gws}{nswmap}. More explicitly, we consider the following
benchmark values:
\beqs\bea &(n,\rs)=(0,0.015)~~&\mbox{resulting
to}~~(\ns,r)=(0.968,0.044)\,,\label{res2}
\\&(n,\rs)=(0.042,0.025)~~&\mbox{resulting
to}~~(\ns,r)=(0.968,0.028)\,.\label{res3}\eea\eeqs

The outputs of our computation is listed in the two rightmost
columns of \Tref{tab}. Since the required $\lm$'s are compatible
with \eqs{lm1}{lm2} for $\nsu=2$, we conclude that the whole
inflationary scenario can be successfully combined with CMSSM. The
$\lm$ values are lower compared to those found in \cref{R2r}.
Moreover, in sharp contrast to that model, all the CMSSM regions
can be consistent with the gravitino limit on $\Trh$ -- see
\Sref{lept1}. Indeed, $\mgr$ as low as $1~\TeV$ become
cosmologically safe, under the assumption of the unstable $\Gr$,
for the $\Trh$ values, necessitated for satisfactory leptogenesis,
as presented in \Tref{tab2}. From the analysis above it is evident
that the solution of the $\mu$ problem in our model becomes a
bridge connecting the high with the low-energy phenomenology.

\renewcommand{\arraystretch}{1.25}
\begin{table}[!t] \bec
\begin{tabular}{|c|c|c|c||c|c|c|}\hline
{\sc CMSSM}&$|A_0|$&$m_0$&$|\mu|
$&$\am$&\multicolumn{2}{|c|}{\sc $\lm (10^{-6})$ for $(\ns,r)$ in}\\
\cline{6-7} {\sc Region }&$(\TeV)$&$(\TeV)$&$(\TeV)$&&\Eref{res2}&
\Eref{res3}\\\hline\hline
$A/H$ Funnel &$9.9244$ &$9.136$&$1.409$&$1.086$ &$0.441$&
$0.6045$\\
$\tilde\tau_1-\chi$ Coannihilation &$1.2271$ &$1.476$&$2.62$& $0.831$&$6.63$& $9.1$\\
$\tilde t_1-\chi$ Coannihilation  &$9.965$ &$4.269$&$4.073$&$2.33$ &$1.27$& $1.74$\\
$\tilde \chi^\pm_1-\chi$ Coannihilation  &$9.2061$ &$9.000$&$0.983$&$1.023$ &$0.332$& $0.454$\\
\hline
\end{tabular}
\end{center}
\caption[]{\sl\small The required $\lm$ values which render our
models compatible with the best-fit points in the CMSSM, as found
in \cref{mssm}, for $m_0=\mgr$, $|A_\ld|=|\aS|=|A_0|$, $K=K_2$ or
$K_3$ with $\nsu=2$ and $(n,\rs)$ given in \eqs{res2}{res3}.}
\label{tab}
\end{table}\renewcommand{\arraystretch}{1.}

\section{Non-Thermal Leptogenesis and Neutrino Masses}\label{pfhi}

We below specify how our inflationary scenario makes a transition
to the radiation dominated era (\Sref{lept0}) and offers an
explanation of the observed BAU (\Sref{lept1}) consistently with
the $\Gr$ constraint and the low energy neutrino data
(\Sref{lept2}). Our results are summarized in \Sref{num}.

\subsection{Inflaton Mass \& Decay}\label{lept0}

The transition to the radiation epoch is controlled by the
inflaton mass and its decay channels. These issues are investigated below
in \Srefs{msnsec} and \ref{decaysec} respectively.

\subsubsection{Mass Spectrum at the SUSY Vacuum} \label{msnsec}

When HI is over, the inflaton continues to roll down towards the
SUSY vacuum, \Eref{vevs}. Soon after, it settles into a phase of
damped oscillations around the minimum of $\Vhi$. The (canonically
normalized) inflaton,
\beq\dphi=\vev{J}\dph\>\>\>\mbox{with}\>\>\> \dph=\phi-M
\>\>\>\mbox{and}\>\>\>\vev{J}=\sqrt{\vev{\kp_+}}\simeq\sqrt{\cm(1-{N\rs})}\label{dphi}
\eeq
acquires mass, at the SUSY vacuum in \Eref{vevs}, which is given
by
\beq \label{msn} \msn=\left\langle\Ve_{\rm
HI,\se\se}\right\rangle^{1/2}= \left\langle \Ve_{\rm
HI,\sg\sg}/J^2\right\rangle^{1/2}\simeq\frac{\ld
M}{\sqrt{2\cm\lf1-{N\rs}\rg}}\,,\eeq
where the last (approximate) equality above is valid only for
$\rs\ll1/N$ -- see \eqs{kpm}{VJe}.  As we see, $\msn$ depends
crucially on $M$ which may be, in principle, a free parameter
acquiring any subplanckian value without disturbing the
inflationary process.  To determine better our models, though,
we prefer to specify $M$ requiring that $\vev{\Phi}$ and
$\vev{\bar\Phi}$ in \Eref{vevs} take the values dictated by the
unification of the MSSM gauge coupling constants, despite the fact
that $U(1)_{B-L}$ gauge symmetry does not disturb this unification
and $M$ could be much lower. In particular, the unification scale
$\Mgut\simeq2\cdot10^{16}~\GeV$ can be identified with $M_{BL}$ --
see \Tref{tab3} -- at the SUSY vacuum in \Eref{vevs}, i.e.,
\beq \label{Mg} {\sqrt{\cm(\vev{\fr}-N\rs)}gM\over
\sqrt{\vev{\fr}}}=\Mgut\>\Rightarrow\>M\simeq{\Mgut}/{g\sqrt{\cm\lf1-{N\rs}\rg}}\eeq
with $g\simeq0.7$ being the value of the GUT gauge coupling and we
take into account that $\vev{\fr}\simeq1$. Upon substitution of
the last expression in \Eref{Mg} into \Eref{msn} we can infer that
$\msn$ remains constant for fixed $n$ and $\rs$ since $\ld/\cm$ is
fixed too -- see \Eref{lan}. Particularly, along the bold solid
line in \sFref{fig1}{b} we obtain
\beqs\bea \label{resmass} &5.8\cdot10^{11}\lesssim
{\msn/\GeV}\lesssim3.6\cdot10^{13}&~~\mbox{for}~~K=K_1;\\
&5.3\cdot10^{10}\lesssim
{\msn/\GeV}\lesssim3.6\cdot10^{13}&~~\mbox{for}~~K=K_2~~\mbox{and}~~K_3\,,
\eea\eeqs
where the lower [upper] bound is obtained for
$(n,\rs)=(-0.121,0.0125)$ [$(n,\rs)=(0.0215,0.499)$ for $K=K_2$
and $K_3$ or $(n,\rs)=(0.0215,0.33)$ for $K=K_1$] -- see
\Eref{res1}. We remark that $\msn$ is heavily affected from the
choice of $K$'s in Eqs.~(\ref{K1}) -- (\ref{K3}) as $\rs$
approaches its lower bound in \sFref{fig1}{a} -- note that this
point is erroneously interpreted in \cref{var}.  For any choice of
$K$ we observe that $\msn$ approaches its value within pure nMI
\cite{nmH} and Starobinsky inflation \cite{R2r, su11} as $\rs$
approaches its maximal value in \Eref{rsmin} -- or as $r$
approaches $0.003$.


\subsubsection{Inflaton Decay}\label{decaysec} The decay of $\dphi$ is processed through the following decay
channels \cite{Idecay}:

\paragraph{(a) Decay channel into {\nsz $\sni$}'s. } The lagrangian which
describes these decay channels arises from the part of the SUGRA
langrangian \cite{nilles} containing two fermions. In particular,
\beqs\bea \nonumber \Lg_{\dphi\to
\sni\sni}&=&-\frac12e^{K/2\mP^2}W_{{\rm HI},N_i^cN^c_i}\sni\sni\
+{\rm
h.c.}=\frac{\ld_{iN^c}}{2}\lf1+\cp\frac{\sg^2}{\mP^2}\rg^{-N/2}\sg\sni\sni\
+{\rm h.c.}\\&=& g_{iN^c} \dphi\ \sni\sni+{\rm
h.c.}\>\>\mbox{with}\>\>g_{iN^c}=\frac{\ld_{iN^c}}{2\vev{J}}\lf1-3\cp
\frac{N}{2}\frac{M^2}{\mP^2}\rg\,, \label{Lnu}\eea where the
masses of $\sni$'s are obtained from the third term of the r.h.s
in \Eref{Whi} as follows
\beq \mrh[i]=\ld_{iN^c}M/f_{0\cal
R}^{N/2}~~~\mbox{with}~~~f_{0\cal R}=1+\cp
M^2/\mP^2\,~~~\mbox{and}~~~\mrh[i]\leq7.1M\,,\label{mrh}\eeq
due to the needed perturbativity of $\ld_{iN^c}$, i.e.,
$\ld_{iN^c}^2/4\pi\leq1$. The result in \Eref{Lnu} can be
extracted, if we perform an expansion for $\mP\to\infty$ and then
another about $\vev{\phi}$. This channel gives rise to the
following decay width
\beq
\GNsn=\frac{1}{16\pi}g_{iN^c}^2\msn\lf{1-4M_{iN^c}^2/\msn^2}\rg^{3/2},
\label{Gpq}\eeq\eeqs
where we take into account that $\dphi$ decays into identical
particles.

\paragraph{(b) Decay channel into {\nsz $\hu$} and {\nsz $\hd$}.}  The lagrangian term which describes the relevant interaction comes from the F-term SUGRA scalar
potential in \Eref{Vsugra}. Namely, we obtain
\beqs\bea\nonumber {\cal L}_{\dphi\to
\hu\hd}&=&-e^{K/\mP^2}K^{SS^*}\left|W_{{\rm
HI},S}\right|^2=-\frac14\ld\lm f_{\cal R}^{-2(n+1)}\lf
\phi^2-M^2\rg \lf H_u^*H_d^*+{\rm h.c.}\rg +\cdots\\ &=&-g_{H}
\msn\dphi \lf H_u^*H_d^*+{\rm h.c.}\rg
+\cdots\>\>\mbox{with}\>\>g_{H}=\frac{\lm}{\sqrt{2}}\lf1-2\cp(n+1)\frac{M^2}{\mP^2}\rg.\label{Lh}\eea
where we take into account \eqs{ndef}{msn}. This interaction gives
rise to the following decay width
\beq \Ghsn=\frac{2}{8\pi}g_{H}^2\msn, \label{Ghh}\eeq\eeqs
where we take into account that $\hu$ and $\hd$ are $SU(2)_{\rm
L}$ doublets. \eqs{lm1}{lm2} facilitate the reduction of $\Ghsn$
to a level which allows for the decay mode into $\sni$'s playing
its important role for nTL.

\paragraph{(c) Three-particle decay channels.} Focusing on the same part of the SUGRA
langrangian \cite{nilles} as in paragraph (a), for a typical
trilinear superpotential term of the form $W_y=yXYZ$ -- cf.
\Eref{wmssm} --, where $y$ is a Yukawa coupling constant, we
obtain the interactions described by
\beqs\bea  \nonumber {\cal L}_{y\dphi} &=& -\frac12e^{K/2\mP^2}\lf
W_{y,YZ}\psi_{Y}\psi_{Z}+W_{y,XZ}\psi_{X}\psi_{Z}+
W_{y,XY}\psi_{X}\psi_{Y}\rg+{\rm h.c.}\, \\ &=&
-g_y\frac\dphi\mP\lf X\psi_{Y}\psi_{Z}+Y\psi_{X}\psi_{Z}+
Z\psi_{X}\psi_{Y}\rg+{\rm
h.c.}\,~~\mbox{with}\>\>g_y=Ny_{3}\cp\frac{M}{\vev{J}\mP},
\label{Lxyz}\eea
where $\psi_X, \psi_{Y}$ and $\psi_{Z}$ are the chiral fermions
associated with the superfields $X, Y$ and $Z$ whose the scalar
components are denoted with the superfield symbol. Working in the
large $\tb$ regime which yields similar $y$'s for the 3rd
generation, we conclude that the interaction above gives rise to
the following 3-body decay width
\beq \Gysn={n_{\rm f}\over512\pi^3}g_y^2{\msn^3\over\mP^2},
\label{Gpq1}\eeq\eeqs
where for the third generation we take $y\simeq(0.4-0.6)$,
computed at the $\msn$ scale, and $n_{\rm f}=14$ for
$\msn<\mrh[3]$ -- summation is taken over $SU(3)_{\rm C}$ and
$SU(2)_{\rm L}$ indices.

Since the decay width of the produced $\sni$ is much larger than
$\Gsn$ the reheating temperature, $\Trh$, is exclusively
determined by the inflaton decay and is given by \cite{quin}
\beq \label{Trh} \Trh=
\left(72\over5\pi^2g_*\right)^{1/4}\sqrt{\Gsn\mP}
\>\>\>\mbox{with}\>\>\>\Gsn=\GNsn+\Ghsn+\Gysn,\eeq
where $g_*\simeq228.75$ counts the effective number of
relativistic degrees of freedom of the MSSM spectrum at the
temperature $T\simeq\Trh$.

\subsection{Lepton-Number and Gravitino Abundances}\label{lept1}

The mechanism of nTL \cite{lept} can be activated by the
out-of-equilibrium decay of the $N^c_{i}$'s produced by the
$\dphi$ decay, via the interactions in \Eref{Lnu}. If
$\Trh\ll\mrh[i]$, the out-of-equilibrium condition \cite{baryo} is
automatically satisfied. Namely, $N^c_{i}$ decay into (fermionic
and bosonic components of) $H_u$ and $L_i$ via the tree-level
couplings derived from the last term in the r.h.s of
Eq.~(\ref{wmssm}). The resulting -- see \Sref{lept2} --
lepton-number asymmetry $\ve_i$ (per $N^c_i$ decay) after
reheating can be partially converted via sphaleron effects into
baryon-number asymmetry. In particular, the $B$ yield can be
computed as
\beq {\small\sf
(a)}\>\>\>Y_B=-0.35Y_L\>\>\>\mbox{with}\>\>\>{\small\sf
(b)}\>\>\>Y_L=2{5\over4}
{\Trh\over\msn}\sum_{i=1}^3{\GNsn\over\Gsn}\ve_i\,.\label{Yb}\eeq
The numerical factor in the r.h.s of \sEref{Yb}{a} comes from the
sphaleron effects, whereas the one ($5/4$) in the r.h.s of
\sEref{Yb}{b} is due to the slightly different calculation
\cite{quin} of $\Trh$ -- cf.~\cref{baryo}.  The validity of the
formulae above requires that the $\dphi$ decay into a pair of
$\sni$'s is kinematically allowed for at least one species of the
$\sni$'s and also that there is no erasure of the produced $Y_L$
due to $N^c_1$ mediated inverse decays and $\Delta L=1$
scatterings \cite{senoguz}. These prerequisites are ensured if we
impose
\beq\label{kin} {\sf \ftn
(a)}\>\>\msn\geq2\mrh[1]\>\>\>\mbox{and}\>\>\>{\sf \ftn
(b)}\>\>\mrh[1]\gtrsim 10\Trh.\eeq
Finally, the interpretation of BAU through nTL dictates
\cite{plcp} at 95\% c.l.
\beq
Y_B=\lf8.64^{+0.15}_{-0.16}\rg\cdot10^{-11}.\label{BAUwmap}\eeq

The $\Trh$'s required for successful nTL must be compatible with
constraints on the $\Gr$ abundance, $\Yg$, at the onset of
\emph{nucleosynthesis} ({\sf\ftn BBN}). Assuming that $\Gr$ is
much heavier than the gauginos of MSSM, $\Yg$ is estimated to be
\cite{brand,kohri}
\beq\label{Ygr} \Yg\simeq 1.9\cdot10^{-22}\Trh/\GeV, \eeq
where we take into account only the thermal $\Gr$ production.
Non-thermal contributions to $\Yg$ \cite{Idecay} are also possible
but strongly dependent on the mechanism of soft SUSY breaking.
Moreover, no precise computation of this contribution exists
within HI adopting the simplest Polonyi model of SUSY breaking
\cite{grNew}. For these reasons, we here adopt the conservative
estimation of $\Yg$ in \Eref{Ygr}. Nonetheless, it is notable that
the non-thermal contribution to $\Yg$ in models with stabilizer
field, as in our case, is significantly suppressed compared to the
thermal one.

On the other hand, $\Yg$  is bounded from above in order to avoid
spoiling the success of the BBN. For the typical case where $\Gr$
decays with a tiny hadronic branching ratio, we have \cite{kohri}
\beq  \label{Ygw} \Yg\lesssim\left\{\bem
%
10^{-15}\hfill \cr
10^{-14}\hfill \cr
10^{-13}\hfill \cr
10^{-12}\hfill \cr \eem
\right.\>\>\>\mbox{for}\>\>\>\mgr\simeq\left\{\bem
0.43~\TeV\hfill \cr
0.69~\TeV\hfill \cr
10.6~\TeV\hfill \cr
13.5~\TeV\hfill \cr \eem
\right.\>\>\>\mbox{implying}\>\>\>\Trh\lesssim5.3\cdot\left\{\bem
%
10^{6}~\GeV\,,\hfill \cr
10^{7}~\GeV\,,\hfill \cr
10^{8}~\GeV\,,\hfill \cr
10^{9}~\GeV\,.\hfill \cr\eem
\right.\eeq
The bounds above can be somehow relaxed in the case of a stable
$\Gr$ -- see e.g. \cref{grlsp}. In a such case, $\Gr$ should be
the LSP and has to be compatible with the data \cite{plcp} on the
CDM abundance in the universe. To activate this scenario we need
lower $\mgr$'s than those obtained in \Sref{secmu2}. As shown from
\Eref{mu1}, this result can be achieved for lower $\mu$'s and/or
larger $\am$'s. Low $\rs$'s, implying large $r$'s, generically
help in this direction too.

Note, finally, that both \eqs{Yb}{Ygr} calculate the correct
values of the $B$ and $\Gr$ abundances provided that no entropy
production occurs for $T<\Trh$. This fact can be achieved if the
Polonyi-like field $z$ decays early enough without provoking a
late episode of secondary reheating.  A subsequent difficulty is
the possible over-abundance of the CDM particles which are
produced by the $z$ decay -- see \cref{olivegr}.

\subsection{Lepton-Number Asymmetry and Neutrino Masses}\label{lept2}

As mentioned above, the decay of $\ssni$, emerging from the
$\dphi$ decay, can generate a lepton asymmetry, $\ve_i$, caused by
the interference between the tree and one-loop decay diagrams,
provided that a CP-violation occurs in $h_{ijN}$'s. The produced
$\ve_i$ can be expressed in terms of the Dirac mass matrix of
$\nu_i$, $m_{\rm D}$, defined in the $\sni$-basis, as follows
\cite{covi}:
\beqs\beq\ve_i ={\sum_{j\neq i}
\im\left[(\mD[]^\dag\mD[])_{ij}^2\right]
\over8\pi\vev{H_u}^2(\mD[]^\dag\mD[])_{ii}}\bigg( F_{\rm S}\lf
x_{ij},y_i,y_j\rg+F_{\rm V}(x_{ij})\bigg), \label{el}\eeq where we
take $\vev{H_u}\simeq174~\GeV$, for large $\tan\beta$ and \beq
x_{ij}={\mrh[j]\over\mrh[i]},\>\>\>F_{\rm V}\lf x\rg=-x\ln\lf1+
x^{-2}\rg~~\mbox{and}~~F_{\rm S}\lf x\rg={-2x\over
x^2-1}\cdot\eeq\eeqs

The involved in \Eref{el} $\mD[]$ can be diagonalized if we define
a basis -- called \emph{weak basis} henceforth -- in which the
lepton Yukawa couplings and the $SU(2)_{\rm L}$ interactions are
diagonal in the space of generations. In particular we have
\beq \label{dD} U^\dag\mD[]U^{c\dag}=d_{\rm
D}=\diag\lf\mD[1],\mD[2], \mD[3]\rg,\eeq where $U$ and $U^c$ are
$3\times3$ unitary matrices which relate $L_i$ and $\sni$ (in the
$\sni$-basis) with the ones $L'_i$ and $\nu^{c\prime}_i$ in the
weak basis as follows
\beq L'= L U\>\>\> \mbox{and}\>\>\>N^{c\prime}=U^c N^c.\eeq
Here, we write LH lepton superfields, i.e. $SU(2)_{\rm L}$ doublet
leptons, as row 3-vectors in family space and RH anti-lepton
superfields, i.e. $SU(2)_{\rm L}$ singlet anti-leptons, as column
3-vectors. Consequently, the combination $\mD[]^\dag\mD[]$
appeared in \Eref{el} turns out to be a function just of $d_{\rm
D}$ and $U^c$. Namely, \beq\mD[]^\dag\mD[]=U^{c\dag} d_{\rm
D}d_{\rm D}U^c. \label{mDD}\eeq

The connection of the leptogenesis scenario with the low energy
neutrino data can be achieved through the seesaw formula, which
gives the light-neutrino mass matrix $m_\nu$ in terms of $\mD[i]$
and $\mrh[i]$. Working in the $\sni$-basis, we have
\beq \label{seesaw} m_\nu= -m_{\rm D}\ d_{N^c}^{-1}\ m_{\rm
D}^{\tr},\eeq where \beq\label{seesaw1} d_{N^c}=
\diag\lf\mrh[1],\mrh[2],\mrh[3]\rg \eeq with
$\mrh[1]\leq\mrh[2]\leq\mrh[3]$ real and positive. Solving
\Eref{dD} w.r.t $\mD[]$ and inserting the resulting expression in
\Eref{seesaw} we extract the mass matrix
\beqs\beq \label{bmn} \bar m_\nu=U^\dag m_\nu U^*=-d_{\rm
D}U^cd_{N^c}^{-1}U^{c\tr}d_{\rm D},\eeq which can be diagonalized
by the unitary PMNS matrix satisfying \beq \bar m_\nu=U_\nu^*\
\diag\lf\mn[1],\mn[2],\mn[3]\rg\ U^\dag_\nu\label{mns1}\eeq and
parameterized as follows
\beq \label{mns2} U_\nu = \mtn{c_{12}c_{13}}{s_{12}c_{13}}{s_{13}
e^{-i\delta}} {-c_{23}s_{12}-s_{23}c_{12}s_{13}
e^{i\delta}}{c_{23}c_{12}-s_{23}s_{12}s_{13}
e^{i\delta}}{s_{23}c_{13}} {s_{23}s_{12}-c_{23}c_{12}s_{13}
e^{i\delta}}{-s_{23}c_{12}-c_{23}s_{12}s_{13}
e^{i\delta}}{c_{23}c_{13}}\cdot \diag\lf
e^{-i\varphi_1/2},e^{-i\varphi_2/2},1\rg, \eeq\eeqs where
$c_{ij}:=\cos \theta_{ij}$, $s_{ij}:=\sin \theta_{ij}$ and
$\delta$, $\varphi_1$ and $\varphi_2$ are the CP-violating Dirac
and Majorana phases.

Following a bottom-up approach, along the lines of \cref{senoguz,
R2r, rob, quad}, we can find $\bar m_\nu$ via \Eref{mns1} adopting
the normal or inverted hierarchical scheme of neutrino masses. In
particular, $\mn[i]$'s can be determined via the relations
\beq \label{mns} \mn[2]=\sqrt{\mn[1]^2+\Delta m^2_{21}}
~~\mbox{and}~~\left\{\bem
\mn[3]=\sqrt{\mn[1]^2+\Delta m^2_{31}}, \hfill& \mbox{for
\emph{normally ordered} ({\sf\ftn NO}) $\mn[]$'s} \cr
\mbox{or} &\cr
\mn[1]=\sqrt{\mn[3]^2+\left|\Delta m^2_{31}\right|},\hfill   &
\mbox{for \emph{invertedly ordered} ({\sf\ftn  IO}) $\mn[]$'s}.
\cr\eem
\right.\eeq where the neutrino mass-squared differences $\Delta
m^2_{21}$ and $\Delta m^2_{31}$ are listed in \Tref{tabn} and
computed by the solar, atmospheric, accelerator and reactor
neutrino experiments. We also arrange there the inputs on the
mixing angles $\theta_{ij}$ and on the CP-violating Dirac phase,
$\delta$, for normal [inverted] neutrino mass hierarchy
\cite{valle} -- see also \cref{lisi}.  Moreover, the sum of
$\mn[i]$'s is bounded from above by the current data \cite{plcp},
as follows \beq \mbox{$\sum_i$} \mn[i]\leq0.23~{\eV}~~\mbox{at
95\% c.l.}\label{mnol}\eeq

\renewcommand{\arraystretch}{1.1}
\begin{table}[!t]
\begin{center}
\begin{tabular}{|c|c|c|}\hline
{\sc Parameter }&\multicolumn{2}{c|}{\sc Best Fit
$\pm1\sigma$}\\\cline{2-3} &{\sc Normal}&{\sc Inverted}
\\\cline{2-3} &\multicolumn{2}{c|}{\sc Hierarchy} \\
\hline\hline
$\Delta
m^2_{21}/10^{-5}\eV^2~~~~$&\multicolumn{2}{c|}{$7.6^{+0.19}_{-0.18}$}\\\cline{2-3}
$\Delta
m^2_{31}/10^{-3}\eV^2$&$2.48^{+0.05}_{-0.07}$&{$2.38^{+0.05}_{-0.06}$}\\\hline
$\sin^2\theta_{12}/0.1$ &
\multicolumn{2}{c|}{$3.23\pm0.16$}\\\cline{2-3}
$\sin^2\theta_{13}/0.01$&$2.26\pm0.12$&$2.29\pm0.12$\\
$\sin^2\theta_{23}/0.1$&$5.67^{+0.32}_{-1.24}$&$5.73^{+0.25}_{-0.39}$\\\hline
$\delta/\pi~~$&$~~1.41^{+0.55}_{-0.4}~~~$&$~~~1.48\pm0.31~~~$\\
\hline
\end{tabular}\end{center}
\caption{\sl\small Low energy experimental neutrino data for
normal or inverted hierarchical neutrino masses. }\label{tabn}
\end{table}
\renewcommand{\arraystretch}{1.}

Taking also $\mD[i]$ as input parameters we can construct the
complex symmetric matrix
\beqs\beq \mathbb{W}=-d_{\rm D}^{-1}\bar m_\nu d_{\rm
D}^{-1}=U^cd_{N^c}U^{c\tr}\label{Wm}\eeq
-- see \Eref{bmn} -- from which we can extract $d_{N^c}$ as
follows \beq d_{N^c}^{-2}=U^{c\dag}\mathbb{W} \mathbb{W}^\dag
U^c\,.\label{WW}\eeq\eeqs Note that $\mathbb{W} \mathbb{W}^\dag$
is a $3\times3$ complex, hermitian matrix and can be diagonalized
numerically  so as to determine the elements of $U^c$ and the
$\mrh[i]$'s. We then compute $\mD[]$ through \Eref{mDD} and the
$\ve_i$'s through \Eref{el}.


\subsection{Results}\label{num}

The success of our inflationary scenario can be judged, if, in
addition to the constraints of \Sref{fhi2}, it can become
consistent with the post-inflationary requirements mentioned in
\Srefs{lept1} and \ref{lept2}. More specifically, the quantities
which have to be confronted with observations are $Y_B$ and $\Yg$
which depend on $\msn$, $\Trh$, $\mrh[i]$ and $\mD[i]$'s  -- see
\eqs{Yb}{Ygr}. As shown in \Eref{msn}, $\msn$ is a function of $n$
and $\rs$ whereas $\Trh$ in \Eref{Trh} depend on $\lm$, $y$ and
the masses of the $\sni$'s into which $\dphi$ decays. Throughout
our computation we fix $y=0.5$ which is a representative value.
Also, when we employ $K=K_1$ and $K_2$ we take $\nb=2$ which
allows for a quite broad available $\lm$ margin. As regards the
$\nu_i$ masses, we follow the bottom-up approach described in
\Sref{lept2}, according to which we find the $\mrh[i]$'s by using
as inputs the $\mD[i]$'s, a reference mass of the $\nu_i$'s --
$\mn[1]$ for NO $\mn[i]$'s, or $\mn[3]$ for IO $\mn[i]$'s --, the
two Majorana phases $\varphi_1$ and $\varphi_2$ of the PMNS
matrix, and the best-fit values, listed in \Tref{tabn}, for the
low energy parameters of neutrino physics. In our numerical code,
we also estimate, following \cref{running}, the RG evolved values
of the latter parameters at the scale of nTL, $\Lambda_L=\msn$, by
considering the MSSM with $\tan\beta\simeq50$ as an effective
theory between $\Lambda_L$ and the soft SUSY breaking scale,
$M_{\rm SUSY}=1.5~\TeV$. We evaluate the $\mrh[i]$'s at
$\Lambda_L$, and we neglect any possible running of the $\mD[i]$'s
and $\mrh[i]$'s. The so obtained $\mrh[i]$'s clearly correspond to
the scale $\Lambda_L$.


\renewcommand{\arraystretch}{1.2}
\begin{table}[!t]
\bec\begin{tabular}{|c||c|c||c|c|c||c|c|}\hline
{\sc Parameters} &  \multicolumn{7}{c|}{\sc Cases}\\\cline{2-8}
&A&B& C & D& E & F&G\\ \cline{2-8} &\multicolumn{2}{c||}{\sc
Normal} & \multicolumn{3}{|c||}{\sc Almost}&
\multicolumn{2}{|c|}{\sc Inverted}
\\& \multicolumn{2}{c||}{\sc Hierarchy}&\multicolumn{3}{|c||}{\sc Degeneracy}&
\multicolumn{2}{|c|}{\sc Hierarchy}\\ \hline  \hline
\multicolumn{8}{|c|}{\sc Low Scale Parameters}\\\hline
$\mn[1]/0.1~\eV$&$0.05$&$0.1$&$0.5$ & $0.7$& $0.7$ & $0.5$&$0.49$\\
$\mn[2]/0.1~\eV$&$0.1$&$0.13$&$0.51$ & $0.7$& $0.7$ & $0.51$&$0.5$\\
$\mn[3]/0.1~\eV$&$0.5$&$0.51$&$0.7$ & $0.86$&$0.5$ &
$0.1$&$0.05$\\\hline
$\sum_i\mn[i]/0.1~\eV$&$0.65$&$0.74$&$1.7$ & $2.3$&$1.9$ &
$1.1$&$1$\\ \hline
$\varphi_1$&$-\pi/8$&$-\pi$&$\pi$ & $\pi/2$&$0$ & $0$&$\pi$\\
$\varphi_2$&$\pi$&$0$ &$\pi/3$& $-\pi$&$-\pi/2$ &
$-\pi/3$&$-\pi/3$\\\hline
\multicolumn{8}{|c|}{\sc Leptogenesis-Scale Parameters}\\\hline
$\mD[1]/0.1~\GeV$&$2$&$2.37$&$10$ & $7.3$&$4$ & $15$&$12$\\
$\mD[2]/\GeV$&$2.2$&$1.3$&$7.5$ & $5$&$9$ & $0.9$&$0.9$\\
$\mD[3]/\GeV$&$100$&$250$&$170$ & $250$&$1.3$ &
$180$&$270$\\\hline
$\mrh[1]/10^{10}~\GeV$&$2.33$&$1.3$&$2.97$ & $0.9$&$0.28$ & $3.11$&$2.93$\\
$\mrh[2]/10^{10}~\GeV$&$7.8$&$4.5$&$92.7$ & $137.6$&$2.4$ & $3.76$&$3.16$\\
$\mrh[3]/10^{14}~\GeV$&$2.9$&$10.4$&$2.3$ &
$1.1$&$9.2\cdot10^{-3}$ & $13.8$&$51.9$\\\hline
\multicolumn{8}{|c|}{\sc Open Decay Channels of the Inflaton,
\dphi, Into $\sni$}\\\hline
$\dphi\ \to$&$\wrhn[1]$&$\wrhn[1]$& $\wrhn[1]$& $\wrhn[1]$&
$\wrhn[1,2]$ & $\wrhn[1,2]$&$\wrhn[1,2]$\\ \hline
$\sum_i\GNsn/\Gsn~(\%)$&$16.5$&$8.2$& $16.9$& $4.5$& $17$ & $22.5$&$28.3$\\
\hline
\multicolumn{8}{|c|}{\sc Resulting $B$-Yield }\\\hline
$10^{11}Y^{(0)}_B$&$9.5$&$9.2$&$6.6$&$9.2$&$10.3$&$6.6$&$9.3$\\
$10^{11}Y_B$&$8.67$&$8.68$&$8.6$ & $8.65$&$8.65$ &
$8.72$&$8.78$\\\hline
\multicolumn{8}{|c|}{\sc Resulting $\Trh$ and $\Gr$-Yield
}\\\hline
$\Trh/10^{7}~\GeV$&$2.8$&$2.7$&$2.83$ & $2.78$&$2.83$ & $2.93$&$3$\\
$10^{15}\Yg$&$5.4$&$5.1$&$5.4$ & $5.3$&$5.4$ &
$5.56$&$5.78$\\\hline
\end{tabular}\eec
\hfill \caption[]{\sl\small  Parameters yielding the correct $Y_B$
for various neutrino mass schemes. We take $K=K_2$ or $K_3$ with
$\nsu=2$, $(n,\rs)$ in \Eref{res3}, $\lm=10^{-6}$ and $y=0.5$.}
\label{tab2}
\end{table}

We start the exposition of our results arranging in \Tref{tab2}
some representative values of the parameters which yield $\Yb$ and
$\Yg$ compatible with \eqs{BAUwmap}{Ygw}, respectively. We set
$\lm=10^{-6}$ in accordance with \eqs{lm1}{lm2}. Also, we select
the $(n,\rs)$ value in \Eref{res3} which ensures central $n$ and
$r$ in \sEref{nswmap}{a} and (\ref{gws}). We obtain
$M=2.39\cdot10^{15}~\GeV$ and $\msn=8.8\cdot10^{10}~\GeV$ for
$K=K_1$ or $M=2.43\cdot10^{15}~\GeV$ and
$\msn=8.6\cdot10^{10}~\GeV$ for $K=K_2$ or $K_3$. Although such
uncertainties from the choice of $K$'s do not cause any essential
alteration of the final outputs, we mention just for definiteness
that we take $K=K_2$ or $K_3$ throughout. We consider NO (cases A
and B), almost degenerate (cases C, D and E) and IO (cases F and
G) $\mn[i]$'s. In all cases, the current limit of \Eref{mnol}  is
safely met -- in the case D this limit is almost saturated. We
observe that with NO or IO $\mn[i]$'s, the resulting $\mrh[1]$ and
$\mrh[2]$ are of the same order of magnitude, whereas these are
more strongly hierarchical with degenerate $\mn[i]$'s. In all
cases, the upper bounds in \Eref{mrh} is preserved thanks to the
third term adopted in the r.h.s of \Eref{Whi} -- cf. \cref{nmH}.
We also remark that $\dphi$ decays mostly into $N_1^c$'s -- see
cases A -- D. From the cases E -- G, where the decay of $\dphi$
into $N_2^c$ is unblocked, we notice that, besides case E, the
channel $\dphi\to N^c_1N^c_1$ yields the dominant contribution to
the calculation $\Yb$ from \Eref{Yb}, since $\what\Gamma_{\dph\to
N^c_1}\geq\what\Gamma_{\dph\to N^c_2}$. We observe, however, that
$\GNsn<\Ghsn$ ($\Gysn$ is constantly negligible) and so the ratios
$\GNsn/\Gsn$ introduce a considerable reduction in the derivation
of $\Yb$. This reduction could have been eluded, if we had adopted
-- as in \crefs{nmH,rob} -- the resolution of the $\mu$ problem
proposed in \cref{rsym} since then, the decay mode in \Eref{Lh}
would have disappeared. This proposal, though, is based on the
introduction of a Peccei-Quinn symmetry, and so the massless
during HI axion generates possibly CDM isocurvature perturbation
which is severely restricted by the \plk\ results \cite{plcp}. In
\Tref{tab2} we also display, for comparison, the $B$ yield with
$(\Yb)$ or without $(\Yb^0)$ taking into account the
renormalization group running of the low energy neutrino data. We
observe that the two results are in most cases close to each other
with the largest discrepancies encountered in cases C, E and F.
Shown are also the values of $\Trh$, the majority of which are
close to $3\cdot10^7~\GeV$, and the corresponding $\Yg$'s, which
are consistent with \Eref{Ygw} for $\mgr\gtrsim1~\TeV$. These
values are in nice agreement with the ones needed for the solution
of the $\mu$ problem of MSSM -- see, e.g., \Fref{fig3} and
\Tref{tab}.

The gauge symmetry considered here does not predict any particular
Yukawa unification pattern and so, the $\mD[i]$'s are free
parameters. For the sake of comparison, however, we mention that
the simplest realization of a SUSY Left-Right [Pati-Salam] GUT
predicts \cite{rob, nick} $h_{iN}=h_{iE}$ [$\mD[i]=m_{iU}$], where
$m_{iU}$ are the masses of the up-type quarks and we ignore any
possible mixing between generations -- these predictions may be
eluded though in more realistic implementations of these models as
in \crefs{nick, rob}. Taking into account the SUSY threshold
corrections \cite{fermionM} in the context of MSSM with universal
gaugino masses and $\tan\beta\simeq50$, these predictions are
translated as follows
\beq \lf m_{1{\rm D}}^0,m_{2{\rm D}}^0,m_{3{\rm D}}^0 \rg\simeq
\left\{\bem
(0.023,4.9,100)~\GeV &\mbox{for a Left-Right GUT},\hfill \cr
(0.0005,0.24,100)~\GeV &\mbox{for a Pati-Salam GUT}.\hfill\cr \eem
\right.\eeq
Comparing these values with those listed in \Tref{tab2}, we remark
that our model is not compatible with any pattern of large
hierarchy between the $\mD[i]$'s, especially in the two lighter
generations, since  and $\mD[1]\gg\mD[1]^0$. On the other hand,
$\mD[2]$ is of the order of $\mD[2]^0$ in cases A -- E whereas
$\mD[3]\simeq m_{3{\rm D}}^0$ only in case A. This arrangement can
be understood, if we take into account that $\mD[1]$ and $\mD[2]$
separately influence the derivation of $\mrh[1]$ and $\mrh[2]$
correspondingly -- see, e.g., \crefs{nmH,senoguz}. Consequently,
the displayed $\mD[1]$'s assist us to obtain the $\ve_1$'s
required by \Eref{BAUwmap}.

\begin{figure}[!t]\vspace*{-.12in}
\hspace*{-.19in}
\begin{minipage}{8in}
\epsfig{file=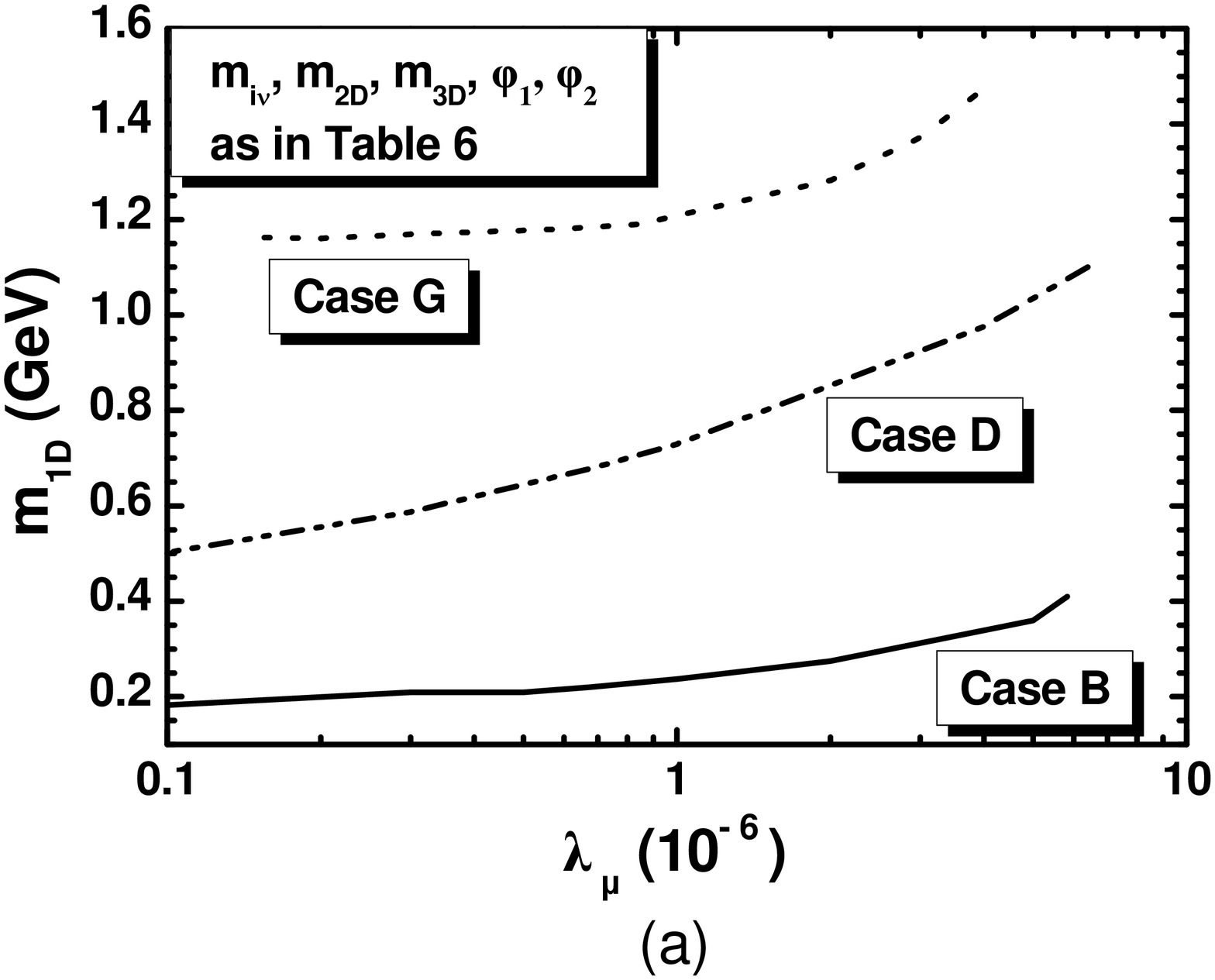,height=3.6in,angle=-90}
\hspace*{-1.2cm}
\epsfig{file=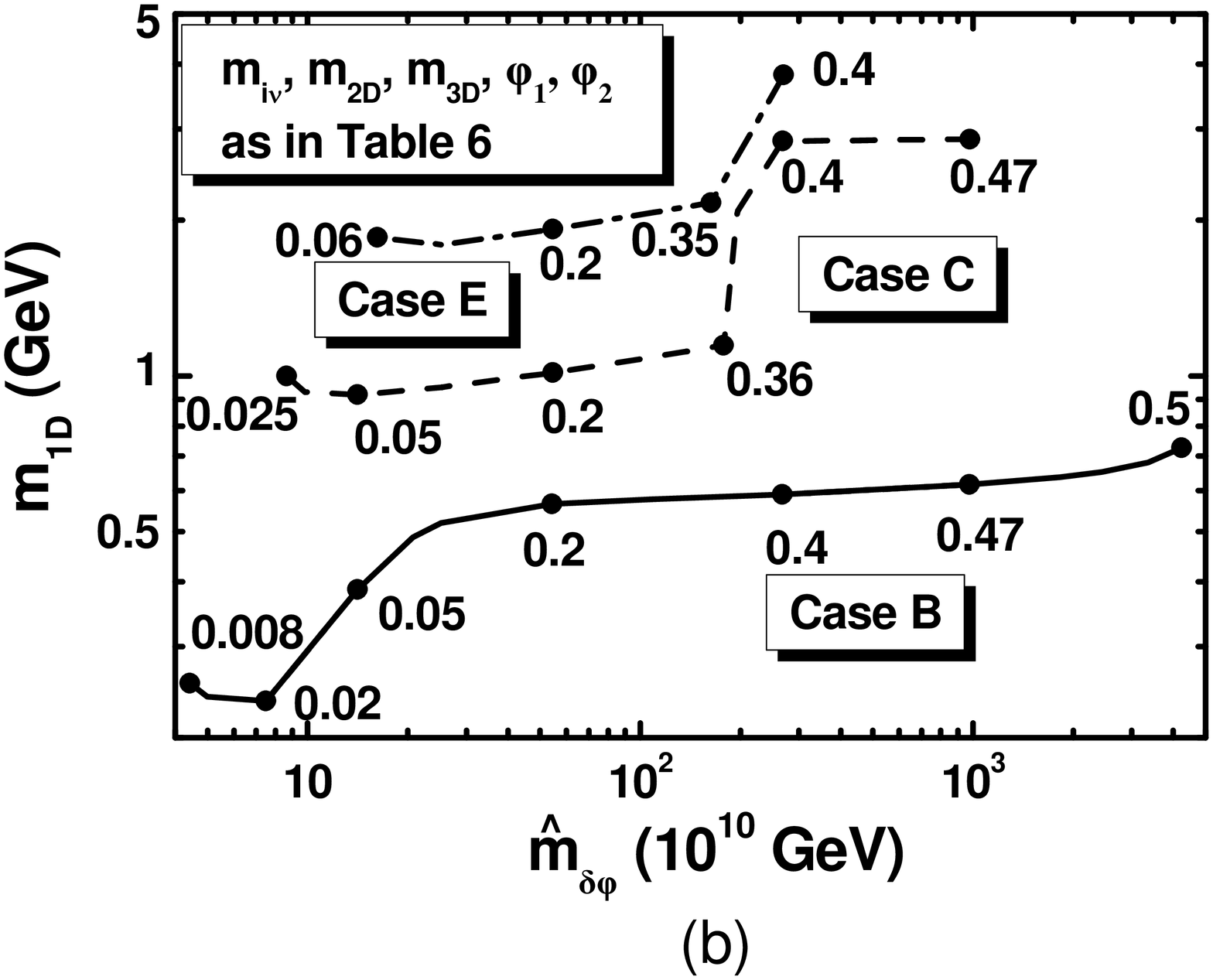,height=3.6in,angle=-90} \hfill
\end{minipage}
\hfill \caption{\sl\small  Contours, yielding the central $Y_B$ in
\Eref{BAUwmap} consistently with the inflationary requirements, in
the {\sffamily\ftn (a)} $\lm-m_{\rm 1D}$ plane for
$(n,\rs)=(0.042,0.025)$; {\sffamily\ftn (b)}
$\widehat{m}_{\dph}-m_{\rm 1D}$ plane for $n=0$ and $\rs$ values
indicated on the curves. We also take $K=K_2$ or $K_3$ with
$\nsu=2$, $y=0.5$ and the values of $m_{i\nu}$, $m_{\rm 2D}$,
$m_{\rm 3D}$, $\varphi_1$, and $\varphi_2$ which correspond to the
cases B (solid line), C (dashed line), D (double dot-dashed line),
E (dot-dashed line), and G (dotted line) of
\Tref{tab2}.}\label{fmD}
\end{figure}

In order to investigate the robustness of the conclusions inferred
from \Tref{tab2}, we examine also how the central value of $Y_B$
in \Eref{BAUwmap} can be achieved by varying $m_{\rm 1D}$ as a
function of $\lm$ and $\msn$ in \sFref{fmD}{a} and {\sf\ftn (b)}
respectively. Since the range of $Y_B$ in \Eref{BAUwmap} is very
narrow, the $95\%$ c.l. width of these contours is negligible. The
convention adopted for these lines is also described in each plot.
In particular, we use solid, dashed, dot-dashed, double dot-dashed
and dotted line when the inputs -- i.e. $\mn[i]$, $\mD[2]$,
$\mD[3]$, $\varphi_1$, and $\varphi_2$ -- correspond to the cases
B, C, E, D, and G of \Tref{tab2}, respectively. In both graphs we
employ $K=K_2$ or $K_3$ with $\nb=2$ and $y=0.5$.

In \sFref{fmD}{a} we fix $(n,\rs)$ to the value used in
\Tref{tab2}. Increasing $\lm$ above its value shown in \Tref{tab2}
the ratio $\GNsn/\Gsn$ gets lower and an increase of $\mrh[1]$ --
and consequently on $\mD[1]$ -- is required to keep $\Yb$ at an
acceptable level. As a byproduct, $\Trh$ and $\Yg$ increase too
and jeopardize the fulfillment of \Eref{Ygw}. Actually, along the
depicted contours in \sFref{fmD}{a}, we obtain
$0.04\leq\Trh/10^8~\GeV\leq1.5$ whereas the resulting $\mrh[1]$'s
[$\mrh[2]$'s] vary in the ranges $(0.8-3)\cdot10^{10}~\GeV$,
$(0.4-2)\cdot10^{10}~\GeV$ and $(2.9-3.1)\cdot10^{10}~\GeV$,
[$(4-6)\cdot10^{10}~\GeV$, $1.3\cdot10^{12}~\GeV$ and
$(3-4)\cdot10^{10}~\GeV$] for the inputs of cases B, D and G
respectively. Finally, $\mrh[3]$ remains close to its values
presented in the corresponding cases of \Tref{tab2}.  At the upper
[lower] termination points of the contours, we obtain $\Yb$ lower
[upper] that the value in \Eref{BAUwmap}.

In \sFref{fmD}{b} we fix $n=0$ and vary $\rs$ in the allowed range
indicated in \sFref{fig1}{a}. Only some segments from that range
fulfill the post-inflationary requirements, despite the fact that
the Majorana phases in Table 6 are selected so as to maximize
somehow the relevant $\msn$ margin. Namely, as inferred by the
numbers indicated on the curves in the $\msn-\mD[1]$ plane, we
find that $\rs$ may vary in the ranges $(0.008-0.499)$,
$(0.025-0.47)$ and $(0.06-0.4)$ for the inputs of cases B, C and E
respectively. The lower limit on these curves comes from the fact
that $\Yb$ is larger than the expectations in \Eref{BAUwmap}. At
the other end, \sEref{kin}{b} is violated and, therefore, washout
effects start becoming significant. At these upper termination
points of the contours, we obtain $\Trh$ of the order $10^9~\GeV$
or $\Yg>10^{-13}$ and so, we expect that the constraint of
\Eref{Ygw} will cut any possible extension of the curves beyond
these termination points that could survive the possible washout
of $Y_L$. As induced by \eqs{msn}{Mg}, $\msn$ increases with $\rs$
and so, an enhancement of $\mrh[1]$'s and similarly of $\mD[1]$'s
is required so that $\Yb$ meets \Eref{BAUwmap}. The enhancement of
$\mD[1]$ becomes sharp until the point at which the decay channel
of $\dphi$ into $N_2^c$'s rendered kinematically allowed.

Compared to the findings of the same analysis in other
inflationary settings \cite{nmH, R2r, quad}, the present scenario
is advantageous since $\msn$ is allowed to reach lower values.
Recall -- see \Sref{lept0} -- that the constant value of $\msn$
obtained in the papers above represents here  the upper bound of
$\msn$ which is approached when $\rs$ tends to its maximal value
in \Eref{rsmin}. In practice, this fact offers us the flexibility
to reduce $\Trh$ and $\Yg$ at a level compatible with $\mgr$
values as light as $1~\TeV$ which are excluded elsewhere. On the
other hand, $\Yb$ increases when $\msn$ decreases and can be kept
in accordance with the expectations due to variation of $\mD[i]$
and $\mrh[i]$. As a bottom line, nTL not only is a realistic
possibility within our models but also it can be comfortably
reconciled with the $\Gr$ constraint.

\section{Conclusions}\label{con}

We investigated the realization of kinetically modified
non-minimal HI (i.e. Higgs Inflation) and nTL (i.e. non-thermal
leptogenesis) in the framework of a model which emerges from MSSM
if we extend its gauge symmetry by a factor $\bl$ and assume that
this symmetry is spontaneously broken at a GUT scale determined by
the running of the three gauge coupling constants. The model is
tied to the super-{} and \Kap s given in Eqs.~(\ref{Whi}) and
(\ref{K1}) -- (\ref{K3}). Prominent in this setting is the role of
a softly broken shift-symmetry whose violation is parameterized by
the quantity $\rs=\cp/\cm$. Combined variation of $\rs$ and $n$ --
defined in \Eref{ndef} -- in the ranges of \Eref{res1} assists in
fitting excellently the present observational data and obtain
$r$'s which may be tested in the near future. Moreover, within our
model, the $\mu$ problem of the MSSM is resolved via a coupling of
the stabilizer field ($S$) to the electroweak higgses, provided
that the relevant coupling constant, $\lm$, is relatively
suppressed. It is gratifying that the derived relation between
$\mu$ and $\mgr$ is compatible with successful low energy
phenomenology of CMSSM. During the reheating phase that follows
HI, the inflaton can decay into $\sni$'s (i.e., right-handed
neutrinos) allowing, thereby for nTL to occur via the subsequent
decay of $\sni$'s. Although other decay channels to the MSSM
particles via non-renormalizable interactions are also activated,
we showed that the generation of the correct $\Yb$, required by
the observations BAU, can be reconciled with the inflationary
constraints, the neutrino oscillation parameters and the $\Gr$
abundance, for masses of the (unstable) $\Gr$ as light as
$1~\TeV$. More specifically, we found that only $N_1^c$ and
$N_2^c$ with masses lower than $1.8\cdot10^{13}~\GeV$ can be
produced by the inflaton decay which leads to a reheating
temperature $\Trh$ as low as $2.7\cdot10^7~\GeV$.


\def\prdn#1#2#3#4{{\sl Phys. Rev. D }{\bf #1}, no. #4, #3 (#2)}
\def\jcapn#1#2#3#4{{\sl J. Cosmol. Astropart.
Phys. }{\bf #1}, no. #4, #3 (#2)}
\def\epjcn#1#2#3#4{{\sl Eur. Phys. J. C }{\bf #1}, no. #4, #3 (#2)}

\end{document}